\documentclass[aps,pre,twocolumn,amsfonts,amsmath,showpacs,a4paper]{revtex4}
\usepackage{color}
\usepackage{enumerate}
\usepackage{graphicx}
\usepackage{times}
\usepackage{float}

\newcommand{\adhoc}{ad hoc}             
\newcommand{\eg}[1]
  {{\it e.g.\/}\ifx#1.\else\expandafter#1\fi}
\newcommand{\eq}[1]{(\ref{#1})}         
\newcommand{\etal}[1]
  {{\it et al.\/}\ifx#1.\else\expandafter#1\fi}
\newcommand{\Fig}[1]{Fig.~\ref{#1}}     
\newcommand{\fig}[1]{fig.~\ref{#1}}     
\newcommand{\ie}[1]
  {{\it i.e.\/}\ifx#1.\else\expandafter#1\fi}
\newcommand{\secn}[1]{Section~\ref{sec:#1}} 
\newcommand{\seclabel}[1]{\label{sec:#1}}  

\newcommand{\dblfigure}[3]
  {\begin{figure*}[tbp]\includegraphics{#1}\caption[]{#2}\label{#3}\end{figure*}}

\renewcommand{\@}{\partial}             
  \newcommand{\<}{\langle}              
\renewcommand{\>}{\rangle}              
\newcommand{\acos}{\arccos}             
\newcommand{\Complex}{\mathbb{C}}       
\newcommand{\Df}[2]{\frac{\d{#1}}{\d{#2}}} 
\renewcommand{\d}{\mathrm{d}}            
\newcommand{\df}[2]{\frac{\@{#1}}{\@{#2}}} 
\newcommand\dtime{\dot }                
\newcommand{\e}{\mathrm{e}}             
\DeclareMathOperator{\Heav}{H}          
\renewcommand{\Im}[1]{{\rm Im}\left(#1\right)}	
\renewcommand{\i}{\mathrm{i}}           
\newcommand{\inner}[2]
  {\left\<#1\, , \,#2\right\>}
\newcommand{\Mx}[1]{
\left[\begin{array}{cccccccc}#1\end{array}\right]}
\newcommand{\mx}[1]{\mathbf{#1}}        
\renewcommand{\Re}[1]{{\rm Re}\left(#1\right)}	
\newcommand{\Real}{\mathbb{R}}          
\newcommand{\T}{^{\mathrm{T}}}          

\newcommand{\Ampl}{A}                   
\newcommand{\Ampres}{\mx{A}}            
\newcommand{\Ampele}{\mx{B}}            
\newcommand{\angled}{\vartheta_0}       
\newcommand{\balp}{\boldsymbol{\alpha}} 
\newcommand{\D}{\mx{D}}                 
\newcommand{\dfdu}{\@_{\u}\f(\U)}       
\newcommand{\dfdp}{\@_{\param}\f}       
\newcommand{\dists}{d}                  
\newcommand{\distd}{l}                  
\newcommand{\F}{F}                      
\newcommand{\Fx}{F_x}                   
\newcommand{\Fy}{F_y}                   
\newcommand{\Fr}{F_r}                   
\newcommand{\Fa}{F_a}                   
\newcommand{\f}{\mx{f}}                 
\newcommand{\g}{\mx{g}}                 
\newcommand{\h}{\mx{h}}                 
\newcommand{\iniphase}{\Theta_0}        
\newcommand{\K}{K}                      
\renewcommand{\L}{\mathcal{L}}          
\newcommand{\Lp}{\L^{+}}                
\renewcommand{\O}{\mathcal{O}}          
\newcommand{\param}{p}                  
\newcommand{\paralp}{\alpha}            
\newcommand{\parbet}{\beta}             
\newcommand{\pargam}{\gamma}            
\newcommand{\para}{a}                   
\newcommand{\parb}{b}                   
\newcommand{\parc}{c}                   
\newcommand{\R}{{\vec R}}               
\newcommand{\Rcore}{R_{\textrm{core}}}  
\newcommand{\Rtip}{R_{\textrm{tip}}}    
\newcommand{\Rrd}{R_{\textrm{rd}}}      
\newcommand{\RF}[1]{\mx{W}^{(#1)}}      
\newcommand{\Ri}{R_{\textrm{in}}}       
\renewcommand{\r}{{\vec r}}             
\newcommand{\rd}{{\vec r_d}}          
\newcommand{\resdrspeed}{p}             
\newcommand{\resdrangvel}{q}            
\newcommand{\Tr}[1]{\mx{V}\ifx#1.\else^{(#1)}\fi} 
\newcommand{\U}{\mx{U}}                 
\renewcommand{\u}{\mx{u}}               
\newcommand{\vF}{{\vec{\F}}}            
\newcommand{\w}[1]{w^{(#1)}}            
\newcommand{\xs}{x_s}                   
\newcommand{\xd}{x_d}                   
\newcommand{\yd}{y_d}                   
\newcommand{\bH}{\textbf{H}}            

\newcommand{\dr}{\Delta\rho}            
\newcommand{\dt}{\Delta t}              
\newcommand{\dx}{\Delta x}              
\newcommand{\Nr}{N_{\rho}}              
\newcommand{\Nt}{N_{\theta}}            
\newcommand{\rp}{\rho_{\max}}           

\begin{document}

\title{Computation of the Drift Velocity of Spiral Waves using Response Functions}

\author{I.V.~Biktasheva}
\affiliation{Department of Computer Science, University of Liverpool, 
     Ashton Building, Ashton Street, Liverpool L69 3BX, UK}
     
\author{D.~Barkley}
\affiliation{Mathematics Institute, University of Warwick, Coventry CV4 7AL, UK}

\author{V. N.~Biktashev}
\affiliation{Department of Mathematical Sciences, University of Liverpool, 
   Mathematical Sciences Building, Peach Street, Liverpool, L69 7ZL, UK}

\author{A.J.~Foulkes}
\affiliation{Department of Computer Science, University of Liverpool, 
     Ashton Building, Ashton Street, Liverpool L69 3BX, UK}

\date{\today}

\begin{abstract}
  Rotating spiral waves are a form of self-organization observed in
  spatially extended systems of physical, chemical, and biological
  nature. In the presence of a small perturbation, the spiral wave's
  centre of rotation and fiducial phase may change over time, i.e. the
  spiral wave drifts. In linear approximation, the velocity of the
  drift is proportional to the convolution of the perturbation with
  the spiral's \emph{Response Functions} (RFs), which are the
  eigenfunctions of the adjoint linearized operator corresponding to
  the critical eigenvalues $\lambda = 0, \pm i\omega$. Here we
  demonstrate that the response functions give quantitatively accurate
  prediction of the drift velocities due to a variety of
  perturbations: a time dependent, periodic perturbation (inducing
  \emph{resonant drift}); a rotational symmetry breaking perturbation
  (inducing \emph{electrophoretic drift}); and a translational
  symmetry breaking perturbation (\emph{inhomogeneity induced drift})
  including drift due to a \emph{gradient, step-wise} and
  \emph{localised} inhomogeneity. We predict the drift velocities
  using the response functions in FitzHugh-Nagumo (FHN) and Barkley
  models, and compare them with the velocities obtained in direct
  numerical simulations. In all cases good quantitative agreement is
  demonstrated.
\end{abstract}


\maketitle

\section{Introduction}
\seclabel{introduction}

Spiral waves are types 
of self-organization observed in
physical \cite{%
  Frisch-etal-1994,%
 Madore-Freedman-1987,%
  Schulman-Seiden-1986%
}, chemical \cite{%
  Zhabotinsky-Zaikin-1971,%
  Jakubith-etal-1990%
}, and biological \cite{%
  Allessie-etal-1973,%
  Gorelova-Bures-1983,%
  Alcantara-Monk-1974,%
  Lechleiter-etal-1991,%
  Carey-etal-1978,%
  Murray-etal-1986%
} systems, where wave propagation is supported by a source of energy
stored in the medium.
The interest in the dynamics of spiral waves has significantly broadened in 
the last decade as the development of experimental techniques has permitted
them to be observed and studied in an ever increasing number of diverse systems
such as %
magnetic films~\cite{Shagalov-1997}, %
liquid crystals~\cite{Oswald-Dequidt-2008}, %
nonlinear optics~\cite{%
  Larionova-etal-2005, %
  Yu-etal-1999%
}, 
novel chemical systems~\cite{Agladze-Steinbock-2000}, %
and in subcellular~\cite{Bretschneider-etal-2009}, %
tissue \cite{Dahlem-Mueller-2003} and %
population biology~\cite{Igoshin-etal-2004}.

In the ideal unperturbed medium, the core of a spiral wave may be
anywhere, depending on initial conditions. However, real systems are
always subject to a perturbation. A typical result of a
symmetry-breaking perturbation is drift of the spiral waves, which has
two components, temporal drift, which is shift of
spiral wave rotation frequency, and spatial drift, that is slow
movement of the spiral's rotation centre. Drift of spiral waves,
particularly the spatial drift, is of great practical interest to
applications. In cardiac tissue, drift of re-entry circuits may be
caused by internal tissue inhomogeneities, or by external
perturbations, such as electrical stimulation. The possibility of control
of arrhythmias by weak electrical stimulation has been a subject of
intensive research for decades.

Understandably, the drift of spiral waves was mostly studied in the BZ
reaction, which is the easiest excitable system for experimental
study, and in the heart tissues and tissue cultures, 
which
represents the most important application area. 
Examples of drift observed
in experiments and numerical simualtions include ``resonant'' drift
caused by (approximately) periodic modulation of medium properties
through external forcing~\cite{Agladze-etal-1987}, constant uniform
electric field that causes electrophoresis of charged ions taking part
in the chemical reactions~\cite{Agladze-Dekepper-1992}, a spatial
gradient of medium properties \cite{%
  Fast-Pertsov-1992,%
  Davidenko-etal-1992,%
  Markus-etal-1992,%
  Luengviriya-etal-2006%
} and pinning (anchoring, trapping) to a localized inhomogeneity
\cite{%
  Nettesheim-etal-1993,
  Pertsov-etal-1993,
  Lim-etal-2006
}. Interaction with a localized inhomogeneity can be considered to be a
particular case of the general phenomenon of vortex pinning to
material defects, ranging from convective microvortex filaments in
nanosecond laser-matter interaction to magnetic flux strings in the Sun's
penumbra~\cite{Lugomer-etal-2007}.

A most intriguing property of spiral waves is that despite being
propagating waves affecting all accessible space, they, or rather
their cores, behave like point-like objects.

Correspondingly, three-dimensional extensions of spiral waves, known
as scroll waves, act as string-like objects. There have been
several \adhoc\ theories of drift of spiral and scroll waves
exploiting incidental features in selected models, \eg~\cite{%
  Yakushevich-1984,%
  Pertsov-Ermakova-1988,%
  Wellner-etal-1999,%
  Hendrey-etal-2000%
}. Our present study is based on an asymptotic theory applicable to
any reaction-diffusion system of equations in which a rigidly rotating
spiral wave solutions exist. The theory was first proposed for autonomous
dynamics of scroll waves for the case of small curvatures and small
twists~\cite{Keener-1988,Biktashev-etal-1994} and then extended to the
drift of spiral waves in response to small
perturbations~\cite{Biktashev-Holden-1995}. In this theory, the
particle-like behaviour of spirals and string-like behaviour of
scrolls corresponds to an effective localization of so called response
functions (RFs, see exact defininition later in this paper).
The localization of RFs is the crucial assumption, which underpins the entire analysis.
Originally~\cite{Biktashev-1989} this property was only a
conjecture based on the phenomenology of spiral waves in experiments and
numerical simulations~\cite{%
  Ermakova-Pertsov-1986,%
  Ermakova-etal-1987,%
  Pertsov-Ermakova-1988,%
  Ermakova-etal-1989%
}. The analytical calculation of the response functions appears to be
infeasible. Numerical calculations in the Barkley model~\cite{Hamm-1997} and
the Complex Ginzburg-Landau equation~\cite{Biktasheva-etal-1998} have
confirmed that indeed they are essentially localized in the vicinity
of the core of the spiral. The asymptotic theory based on the response
functions has been successfully used to quantitatively predict drift
of spirals, for resonant drift and drift due to parametric
inhomogeneity in the CGLE~\cite{%
  Biktasheva-etal-1999,%
  Biktasheva-2000,%
  Biktasheva-Biktashev-2003%
} and for drift in response to a uniform electric field in Barkley
model~\cite{Henry-Hakim-2002}. Despite this success, so far the
asymptotic theory has not become a generally used tool for the prediction
of spiral wave drift. This is partly due to difficulties in the numerical
calculation of the response functions. In our recent
publication~\cite{Biktasheva-etal-2009} we have presented an efficient
numerical method of calculating response functions in an arbitrary
model with differentiable right-hand sides. The complexity of
calculating response functions with this method is similar to the
complexity of calculating spiral wave solutions themselves. In the
present paper, we describe the application of the asymptotic theory using
the response functions for the prediction of several types of drift and show
how it works for two of the most popular generic excitable models, the
FitzHugh-Nagumo
system~\cite{FitzHugh-1961,Nagumo-etal-1962,Winfree-1991}, and the
Barkley system~\cite{Barkley-1991}. We demonstrate that predictions of
the asymptotic theory are in good quantitative agreement with direct
numerical simulations. In addition, we demonstrate that the response functions
are capable of predicting nontrivial qualitative phenomena, such as
attachment of spiral waves to stepwise inhomogeneity and orbital
movement around a localized inhomogeneity.

The structure of the paper is as follows. In \secn{theory}, we briefly
recapitulate the asymptotic theory of the drift of spiral waves in
response to small perturbation and present explicit expressions for
drift parameters in terms of the spiral wave's response functions for
several sorts of drift. In \secn{methods}, we describe the numerical
methods used for calculating the response functions, for direct
numerical simulations, and for processing of the results. The results are
described in \secn{results}. We conclude the paper by discussion of
the results and their implications in \secn{discussion}.

\section{Theory}
\seclabel{theory}

\subsection{General}

We consider reaction-diffusion partial differential equations,
\begin{equation}
\@_t\u = \f(\u) + \D \nabla^2 \u, \quad 
\u,\f\in\Real^\ell,\; 
\D\in\Real^{\ell\times\ell},\;
\ell\ge2,					\label{RDS}
\end{equation}
where $\u(\r,t)=(u_1,\dots u_{\ell})\T$ 
is a column-vector of the reagent concentrations,
$\f(\u)=(f_1,\dots f_{\ell})\T$ is a column-vector of the reaction rates, 
$\D$ is the matrix of diffusion coefficients, and  
$\r\in\Real^2$ is the vector of coordinates on the plane.

A rigidly clockwise rotating spiral wave solution to the system \eq{RDS} has the form 
\begin{equation}
\U=\U(\rho (\r-\R),\vartheta (\r-\R) + \omega t - \Phi) ,
                                      \label{SW}
\end{equation}
where $\R=(X,Y)\T$ is the center of rotation, $\Phi$ is the initial rotation phase, and 
$\rho(\r-\R),\vartheta(\r-\R)$ are polar coordinates centered at $\R$.
For a steady, rigidly rotating
spiral, $\R$ and $\Phi$ are constants. The system of reference 
co-rotating with the spiral's  initial phase and angular velocity $\omega$ around the 
spiral's center of rotation is called the system of reference of the 
spiral. In this system of 
reference, the polar angle is given by $\theta =
\vartheta + \omega t - \Phi$, with $\R=0$ and $ \Phi=0$. In this frame, the spiral 
wave solution $\U(\rho,\theta)$ does not depend on time and satisfies the
equation 
\begin{equation}
\f(\U) - \omega \U_\theta + \D \nabla^2 \U = 0 , 	\label{SW-own}
\end{equation}
where the unknowns are the field $\U(\rho,\theta)$ and the scalar $\omega$.

In a slightly perturbed problem
\begin{equation}
\@_t\u = \f(\u) + \D \nabla^2 \u + \epsilon \h, \quad 
		\h\in\Real^\ell, \quad |\epsilon| \ll 1,
                                       \label{RDS_pert}
\end{equation}
where $\epsilon \h(\u,\r,t)$ is some small perturbation, spiral waves
may drift, \ie\ change rotational phase and/or center location. Then,
the center of rotation and the initial phase are no longer constants
but become functions of time, $\R=\R(t)$ and $\Phi=\Phi(t)$. In the
co-rotating system of reference, time dependence will take form of a
phase depending on time $\phi(t) = \omega t - \Phi(t)$.

Thus, we consider three systems of reference:
\begin{description}
\item[1. laboratory,]  $\left( \r, \; t \right)$ ;
\item[2. co-moving,]  
  $(\rho,\vartheta,t)$, where
  $(\rho,\vartheta)=\left(\rho(\r-\R), \; \vartheta(\r-\R)\right)$
  is the polar coordinate system centered at $\R$;
\item[3. co-rotating,] $\left( \rho, \; \theta, \; \phi\right)$, where
$\theta = \vartheta(\r-\R) + \phi(t)$ is the polar angle, and
$\phi = \omega t - \Phi(t)$ is the rotational phase, replacing time.
\end{description}
We shall look for a solution to \eq{RDS_pert} in the form of a
slightly perturbed steady spiral wave solution
\begin{equation*}
\tilde \U(\rho,\theta,\phi)=\U(\rho,\theta)+ \epsilon \g(\rho,\theta,\phi), 
                                       \label{U_pert}
\end{equation*}
where $\quad \g\in\Real^\ell, \quad 0 < \epsilon \ll 1$. 

Then, assuming that 
\[
  \dtime{\R}, \; \dtime{\Phi} = \O (\epsilon),
\]
at leading order in $\epsilon$, the solution perturbation $\g$ will
satisfy the linearized system

\begin{equation}
( \omega \@_\phi - \L ) \g = \bH(\U,\rho,\theta,\phi),
                       \label{g_eqn_1}             
\end{equation}
where 
\[
\bH(\U,\rho,\theta,\phi) = 
    \tilde\h(\U,\rho,\theta,\phi) 
    - \frac{1}{\epsilon}\bigg[\frac{\@\U}{\@\R} \dtime{\R} 
    - \@_\theta\U \dtime{\Phi} \bigg], 
\]
where $\tilde\h(\U,\rho,\theta,\phi)$
  is the perturbation $\h(\u,\r,t)$,
  considered 
  in the co-rotating frame of reference.

The linearized operator
\begin{equation}
\L = \D\nabla^2 - \omega\@_\theta + \dfdu,        \label{L}
\end{equation}
has critical ($\Re{\lambda}=0$) eigenvalues
\begin{equation}
  \L \Tr{n} = \lambda_n \Tr{n}, 
  \qquad \lambda_n=\i n\omega, 
  \quad n=0,\pm1,
                                                  \label{lambdas}
\end{equation}
which correspond to eigenfunctions related to equivariance of \eq{RDS}
with respect to translations and rotations, \ie\ ``Goldstone modes'' (GMs)
\begin{eqnarray}
\Tr{0} &=&  
  - \@_\theta \U(\rho,\theta), \nonumber\\
\Tr{\pm1} &=& 
  -\frac12 \e^{\mp\i\theta} \left(
    \@_\rho\mp\i\rho^{-1}\@_\theta
  \right) \U(\rho,\theta) .  \label{Goldstone} 
\end{eqnarray}
In this paper we do not consider perturbations $\h(\u,\r,t)$ that
depend on $t$ other than $2\pi/\omega$-periodically (for a more
general version of the theory free from this assumption see
\cite{Biktashev-Holden-1995,Biktasheva-Biktashev-2003}). Then
$\tilde\h(\U,\rho,\theta,\phi)$ is a $2\pi$-periodic function in
$\phi$, and we look for solutions $\g(\rho,\theta,\phi)$ to equation
\eq{g_eqn_1} with the same periodicity. A solvability condition
leads to the following system of equations for the drift velocities,
\begin{eqnarray*}
\dtime{\Phi} &=& \epsilon \int_{0}^{2\pi} 
    \inner{\RF{0}}{\tilde \h(\U,\rho,\theta,\phi) }
    \frac{\d\phi}{2\pi} +\O(\epsilon^2),   \nonumber \\
\dtime{R} &=& \epsilon \int_{0}^{2\pi}e^{-i\phi} 
    \inner{\RF{1}}{\tilde \h(\U,\rho,\theta,\phi) }
    \frac{\d\phi}{2\pi}+\O(\epsilon^2), \nonumber 
\end{eqnarray*}
where $R=X+\i\,Y$ is the complex coordinate of the instant spiral
centre, the inner product $\inner{\cdot}{\cdot}$ stands for the scalar
product in functional space
\[
  \inner{\mx{w}}{\mx{v}} = \int\limits_{\Real^2}
  \mx{w}^+(\r) \, \mx{v}(\r) \,\d^2\r ,
\]
and the kernels $\RF{n}(\rho,\theta)$, $n = 0, \pm 1$, are the
response functions, that is the critical eigenfunctions
\begin{equation}
  \Lp \RF{n} = \mu_n \RF{n},
  \qquad \mu_n=-\i n\omega,
  \quad n=0,\pm1,
                                                  \label{mus}
\end{equation}
of the adjoint operator $\Lp$,
\begin{equation}
  \Lp = \D\nabla^2 + \omega\@_\theta + \left(\dfdu\right)\T ,  
                                                  \label{Lp}
\end{equation}
chosen to be biorthogonal 
\begin{equation}
\inner{ \RF{j} }{ \Tr{k} }=\delta_{j,k} ,  		\label{norm}
\end{equation}
to the Goldstone modes \eq{Goldstone}. 

The drift velocities can be written as  
(henceforth we shall drop the
  $\O(\epsilon^2)$ terms)
\begin{equation}
\dtime\Phi = \epsilon \F_0(\R,\Phi), \quad
\dtime\R   = \epsilon \vF_{1}(\R,\Phi),                     \label{ptb}
\end{equation}
where the ``forces'' 
$\F_0$ and $\vF_1=\left(\Re{F_1},\Im{F_1}\right)\T$   
are defined by
\begin{align}
\F_n(\R,\Phi) = \inner{\RF{n}\left(\rho,\theta\right)}{\balp_n (\rho, \theta; \R, \Phi)} , 
\nonumber\\
n=0,1, \label{forces} 
\end{align}
and
\begin{equation}
\balp_n (\rho, \theta; \R, \Phi) = 
	\int_{0}^{2\pi} e^{-in\phi} \; \tilde \h(\U,\rho,\theta,\phi) \, \frac{\d\phi}{2\pi}. 
  					\label{alphan}
\end{equation}
In the above formulae, the dependence on $(\R,\Phi)$ is explicitly included
to emphasize that the response functions depend on coordinates
$(\rho,\theta)$ in the corotating frame of reference whereas the
perturbations are typically defined in the laboratory frame of
reference, and the two systems of references are related via $\R$ and $\Phi$.

Below we show how the forces \eq{forces}, determining the velocity of
the drifting spiral wave subject to a variety of perturbations, can be
calculated using the computed response functions $\RF{n}$. We also
compare the quantitative analytical prediction of drift velocities with
the results of direct simulations.

\subsection{Resonant Drift}

Let us consider a spiral wave drifting due to the perturbation 
\begin{equation}
  \h(\u,\r,t) = \Ampres \cos(\omega t),           \label{hres}
\end{equation}
where $\Ampres\in\Real^{\ell}$ is a constant vector. 
In the co-rotating frame the perturbation \eq{hres} will be
\begin{equation}
\tilde \h = \Ampres \cos\left(\phi+\Phi\right)   \label{hres_2}
\end{equation}

Substitution of \eq{hres_2} into \eq{alphan} gives
\[
  \balp_0 = 0, \quad \balp_1 = \frac{\Ampres}{2}e^{i\Phi},
\]
and, by \eq{forces},
\begin{equation}
\label{eqn:fb1_res}
\F_0 = 0, \quad 
\F_1 = \frac{1}{2}e^{i\Phi}\inner{\RF{1}(\rho,\theta)}{\Ampres}.
\end{equation}
Hence the \emph{speed} of the resonant drift of the spiral is
\begin{equation}
  \left|\dtime{R} \right| =
  \frac{1}{2}\left|\epsilon\inner{\RF{1}}{\Ampres}\right|,
                                                  \label{res-speed}
\end{equation}
whereas its \emph{direction} is constant and arbitrary,
\begin{equation}
   \arg\left(\dtime{R}\right) =
   \arg\left(\inner{\RF{1}}{\Ampres}\right) + \Phi, \qquad
   \dtime{\Phi}=0, 
                                                  \label{res-direction}
\end{equation}
as it is determined by the
inial phase of the spiral $\Phi$, or, rather, by the phase difference
between the spiral and the perturbation, \eq{res-direction} is
only valid in the asymptotic sense, and a more accurate formulation is
\begin{equation}
  \dtime{\Phi}=\O(\epsilon^2).                    \label{res-direction-2}
\end{equation}
Hence, at finite $\epsilon$ the resonance is expected to be imprecise,
and a typical trajectory of a resonantly drifting spiral is a circle
of radius $\Rrd=|\dtime{R}|/|\dtime{\Phi}|=\O(\epsilon^{-1})$.

\subsection{Electrophoretic Drift}

Here we consider an anisotropic perturbation which breaks rotational symmetry
\begin{equation}
\h(\r) = \Ampele \df{\U}{x} \label{helec}
\end{equation}
where $\Ampele\in\Real^{\ell\times\ell}$ is a constant matrix. This
perturbation corresponds to action of an external electric field on a
chemical reaction where some of the species are electrically
charged. In this case matrix $\Ampele$ is diagonal and its nonzero
elements represent motilities of the ions of the reaction species. The
same sort of perturbation appears in the asymptotic dynamics of scroll
waves~\cite{Keener-1988,Biktashev-etal-1994}, where $\Ampele=\D$.

In the co-rotating system of reference, the perturbation \eq{helec}
can be written using the Goldstone modes \eq{Goldstone}, as
\begin{equation}
\tilde \h(\U,\rho,\theta,\phi) =
  - \Ampele \big( \Tr{-1} e^{-i\phi} + \Tr{1} e^{i\phi} \big) ,  \label{h_electr}
\end{equation}
which, by substituting into \eq{alphan}, gives 
\begin{equation}
\balp_n (\rho, \theta) 
  = - \Ampele \int_{0}^{2\pi}e^{-in\phi} \big(\Tr{-1} e^{-i\phi} +
  \Tr{1} e^{i\phi} \big)  \frac{\d\phi}{2\pi} 
  \label{alphan_electr}
\end{equation}
Thus, $\balp_0=0$, $\balp_1(\rho,\theta) = -\Ampele\Tr{1}$,
which following \eq{ptb} and \eq{forces} gives the velocity of the
electrophoretic drift
\begin{equation}
\dtime{\R} = - \epsilon \inner{\RF{1}(\rho,\theta)}{\Ampele \Tr{1}(\rho, \theta)} 
  					\label{forces_electr}
\end{equation}
which remains constant in time. 

\subsection{Inhomogeneity induced Drift}

\subsubsection{General}

We now consider the case when the reaction kinetics $\f$ in
\eq{RDS} depend on a parameter $\param$, and the value of this
parameter varies slightly in space,
\begin{equation}
  \f = \f(\u,\param), \qquad \param = \param(\r) = \param_0 + \epsilon\param_1(\r). \label{inhom}
\end{equation}
Substitution of \eq{inhom} into \eq{RDS} gives, to the first order in $\epsilon$,
\begin{equation*}
\@_t \u = \D\nabla^2\u+\f(\u, \param_0)+\epsilon \param_1(\r) \dfdp(\u,\param_0),
\end{equation*}
with the perturbation in the laboratory frame of reference
\begin{equation}
\h(\u,\r,t) = \dfdp(\u,\param_0) \, \param_1(\r) .     \label{hinhom}
\end{equation}

Substitution of \eq{hinhom} into \eq{alphan} gives
\begin{equation}
\balp_n(\rho,\theta) = 
  \@_p\f(\U(\rho,\theta),\param_0) \, 
  \e^{-\i n\theta}
  \K_n(\rho),   \label{alpha_inh}
\end{equation}
where 
\begin{equation}
\K_n(\rho) = \int_{0}^{2\pi} \e^{\i n\vartheta} 
  \; \tilde\param_1(\rho,\vartheta) \, \frac{\d\vartheta}{2\pi},
                    \label{Kn_inh}
\end{equation}
and $\tilde\param_1(\rho,\vartheta)$ is the parameter perturbation considered
in the co-moving frame of reference. The final equations for the drift
velocities can then be written in the form
\begin{eqnarray}
\dtime\Phi &=&
  \epsilon \int\limits_{0}^{2\pi} \int\limits_{0}^{\infty}
  \w{0}(\rho,\theta)
  \K_0(\rho) 
  \,\rho\,\d\rho\,\d\theta, \\
\dtime{R}   &=& 
  \epsilon \int\limits_{0}^{2\pi}  \int\limits_{0}^{\infty} 
  \w{1}(\rho,\theta)
  \e^{-\i\theta} 
  \K_1(\rho) 
  \,\rho\,\d\rho\,\d\theta ,
\end{eqnarray}
where for brevity we introduce
\begin{equation}
  \w{n}(\rho,\theta) =
  \left[{\RF{n}(\rho,\theta)}\right]^+\dfdp(\rho,\theta;\param_0) .
\end{equation}

\subsubsection{Linear Gradient}

Let $\param_1$ vary linearly in a sufficiently large region containing the
spiral tip and its subsequent drift trajectory. Specifically we consider
$\param_1=x-x_0$, where the $x$-coordinate of the trajectory remains near
$x_0$.
In the co-moving reference frame, the linear gradient perturbation will be 
\begin{equation}
\tilde p_ 1 = X - x_0 + \rho\cos(\vartheta) . \label{hlinear}
\end{equation}
Substituting \eq{hlinear} into \eq{Kn_inh} gives 
\begin{equation*}
\K_n(\rho) = (X - x_0) \delta_{n,0} \; 
   + \; \frac{1}{2} \rho
   \left( \delta_{n,1} + \delta_{n,-1} \right) .     \nonumber 
\end{equation*}
Then, 
by \eq{alpha_inh},
\eq{ptb} and \eq{forces}, the velocity
of the drift due to gradient of a model parameter will be
\begin{eqnarray}
  \dtime{\Phi} &=& 
  \epsilon (X-x_0)
  \int\limits_{0}^{2\pi} \int\limits_{0}^{\infty}
  \w{0}(\rho,\theta)
  \,\rho\,\d\rho\,\d\theta, 
  \nonumber \\
 \dtime{R} &=& 
 \frac{\epsilon}{2}
 \int\limits_{0}^{2\pi} \int\limits_{0}^{\infty}
 \w{1}(\rho,\theta)
  \e^{-\i\theta} 
 \,\rho^2\,\d\rho\,\d\theta .
                                                  \label{speed_gradient}
\end{eqnarray}

An important feature of equations \eq{speed_gradient} is that the
first of them depends on $X$ while the second does not. The dependence
on $X$ means that the drift velocity changes during the drift,
unless the drift proceeds precisely along the $y$-axis. As it happens,
at first order in $\epsilon$, only the temporal drift, that is the
correction to the frequency, shows this dependence.  Namely, the first
of equations \eq{speed_gradient} shows that the instant rotation
frequency corresponds to the parameter value at the current centre
of rotation, $\param=\param_0+\epsilon\param_1 = \param_0+\epsilon
(X-x_0)$.  The spatial drift, described by the second of equations
\eq{speed_gradient}, does not depend on $X$. That means that while the
drift proceeds, its speed and direction remain the same, at least
at
the asymptotic order considered. This is an important
observation, firstly, because it allows us to treat linear gradient
induced drift in the same way as the electrophoretic drift, \ie\
expecting drift along a straight line, and secondly, that unlike
electrophoretic drift, the assumption is inherently limited to such
$X$ that $\epsilon(X-x_0)$ remains sufficiently small.

\subsubsection{Step inhomogeneity} 
Here we consider a step perturbation located at $x=\xs$, 
\[
p_1(x) = \Heav(x-\xs) ,
\]
where $\Heav()$ denotes the Heaviside unit step function. 
In the co-moving frame of reference we have
\begin{equation}
\tilde p_1(\rho,\vartheta) = \Heav \left( X + \rho\cos(\vartheta)-\xs \right).
                     \label{hstep_1}
\end{equation}
Substitution of \eq{hstep_1} into \eq{Kn_inh} gives 
\begin{equation*}
\K_n = \int_{0}^{2\pi}\cos(n\vartheta) \, 
  \Heav\left( \cos(\vartheta) - \frac{\xs-X}{\rho} \right)
  \, \frac{\d\vartheta}{2\pi} .
\end{equation*}

We consider three intervals for $\frac{\xs - X}{\rho}$. 
\begin{enumerate}[(1)]
\item $\rho < \left|\xs - X \right|, \xs > X$. Then 
      $\Heav\left( \cos(\vartheta) - \frac{\xs - X}{\rho} \right) = 0$, 
      therefore $\K_0=\K_1=0$.
\item $\rho < \left|\xs - X \right|, \xs < X$. Then
      $\Heav\left( \cos(\vartheta)-\frac{\xs - X}{\rho} \right)=1$,
      therefore  $\K_0=1$, $\K_1=0$.
\item $\rho \ge \left|\xs - X \right|$. 
      Then, for $\vartheta_0=\acos\left( \frac{\xs - X}{\rho} \right)$,
\[
  \Heav\left(\cos(\vartheta) - \frac{\xs - X}{\rho}\right) = \left\{\begin{array}{lcl} 
      1 &,& \vartheta\in[-\vartheta_0,\vartheta_0] \\
      0 &,& \textrm{otherwise.}
    \end{array}\right.
\]
Thus,
\begin{eqnarray}
\K_0 &=& \frac{1}{\pi} \acos\left(\frac{\xs - X}{\rho}\right) , \label{step-K0} \\
\K_1 &=& \frac{1}{\pi} \sqrt{1-\left(\frac{\xs - X}{\rho}\right)^2} . \label{step-K1}
\end{eqnarray}
\end{enumerate}
Substituting the above $\K_n$ for the three intervals into \eq{forces}
and \eq{ptb}, we get the velocities of the drift due to a step-wise
inhomogeneity of a model parameter in the form
\begin{widetext}
\begin{eqnarray}
\dtime{R} &=&
  \frac{\epsilon}{\pi} \;
  \int\limits_0^{2\pi} \int\limits_{\left|\xs - X\right|}^\infty \; 
  \w{1}(\rho,\theta)
 \e^{-\i\theta} \; 
  \sqrt{1-\left(\frac{\xs - X}{\rho}\right)^2} \; 
  \rho\,\d\rho\,\d\theta ,
  \label{step-velocity} \\
\dtime{\Phi} &=& 
  \frac{\epsilon}{\pi} \;
  \int\limits_0^{2\pi} \int\limits_{\left|\xs - X\right|}^\infty \; 
  \w{0}(\rho,\theta)
 \acos\left(\frac{\xs - X}{\rho}\right) \;
  \rho\,\d\rho\,\d\theta
  \;+\; 
  \epsilon \Heav(X-\xs)
  \int\limits_0^{2\pi} \int\limits_0^{\left|\xs - X\right|} \;
  \w{0}(\rho,\theta)
  \rho\,\d\rho\,\d\theta .
  \label{step-freqshift}
\end{eqnarray}
\end{widetext}
Note that both $\dtime{R}$ and $\dtime{\Phi}$ are functions of the
current $x$-coordinate of the spiral with respect to the step, $d=X-\xs$,
and $\dtime{R}$ is an even function of this coordinate.

\subsubsection{Disk-shaped inhomogeneity}

We now consider an inhomogeneity which is unity within a disc 
of radius $\Ri$ centered at $(\xd ,\yd)$, and which is zero outside the disc.
Thus we have
\[
  \tilde\param_1(\r)=\Heav\left(\Ri^2-(x-\xd)^2-(y-\yd)^2\right) .
\]
Then calculations, similar to those for a stepwise inhomogeneity, lead to 
\begin{equation}
\K_0 = \frac{1}{\pi}
  \acos\left(\frac{\rho^2 + \distd^2 - \Ri^2}{2\distd\rho} \right) ,
  					          \label{disk-K0} 
\end{equation}
\begin{equation}
\K_1 = \frac{\e^{\i\angled}}{\pi}
  \sqrt{1 - \left(\frac{\rho^2 + \distd^2 - \Ri^2 }{2\distd\rho}\right)^2} ,
				                  \label{disk-K1}
\end{equation}
where $\distd$ and $\angled$ designate the distance and the direction
from the current centre of the spiral to the centre of the
inhomogeneity, \ie\ $\xd=X+\distd\cos\angled$,
$\yd=Y+\distd\sin\angled$.

This leads to the equations for the drift velocities in the form 
\begin{widetext}
\begin{eqnarray}
\dtime{R} &=&
	\frac{\epsilon }{\pi} \e^{\i\angled}
        \int\limits_{0}^{2\pi} \int\limits_{\left| \distd-\Ri \right|}^{\distd + \Ri}
        \w{1}(\rho,\theta)
        \e^{-i\theta} \sqrt{1 - \left(\frac{\rho^2 + \distd^2 - \Ri^2}{2\distd\rho} \right)^2} \;
        \,\rho\,\d\rho\,\d\theta ,
				\label{disk-velocity} \\
\dtime \Phi &=& 
	\frac{\epsilon}{\pi}
        \int\limits_{0}^{2\pi} \int\limits_{\left| \distd-\Ri \right|}^{\distd + \Ri}
        \w{0}(\rho,\theta)
       \acos\left( \frac{\rho^2 + \distd^2 - \Ri^2}{2\distd\rho} \right) \;
        \rho\,\d\rho\,\d\theta , \label{disk-freqshift}
\end{eqnarray}
\end{widetext}
It is straightforward to verify that if $\angled=0$, $\xd=\xs+\Ri$
and $\Ri\to\infty$, that is when the disk is so large it turns into a
half-plane at $x>\xs$, then expressions \eq{disk-K0} and \eq{disk-K1}
tend to expressions \eq{step-K0} and \eq{step-K1} respectively, as
should be expected. Another interesting limit is $\Ri\to0$, in which
we get
\begin{eqnarray}
\dtime \Phi &\approx& \epsilon \pi\Ri^2 
        \int\limits_{0}^{2\pi} 
        \w{0}(l,\theta)
       \frac{\d\theta}{2\pi} \;
        \left(1+\O\left(\frac{\Ri}{l}\right)\right), \nonumber \\
\dtime{R} &\approx&
	\epsilon \pi\Ri^2 \e^{\i\angled}
        \int\limits_{0}^{2\pi} 
        \w{1}(l,\theta)
        \e^{-i\theta} \frac{\d\theta}{2\pi} \; 
        \left(1+\O\left(\frac{\Ri}{l}\right)\right), \nonumber
\end{eqnarray}
in accordance with the case of a pointwise, $\delta$-function
inhomogeneity considered in \cite{OMS}.

\section{Methods}
\seclabel{methods}

\subsection{Models}

We have considered two different kinetic models, both
two-component, $\ell=2$, with one nonzero diffusion coefficient, 
$\D=\Mx{1&0\\0&0}$. We designate
$\u=(u,v)\T$, $\f=(f,g)\T$ for convenience. 
The kinetics FitzHugh-Nagumo system was chosen in
\textcite{Winfree-1991} notation,
\begin{align}
  f(u,v) &= \paralp^{-1}(u-u^3/3-v),                         \nonumber\\
  g(u,v) &= \paralp \, (  u + \parbet - \pargam v ),            \nonumber
\end{align}
with parameter values $\paralp=0.3$, $\parbet=0.68$, $\pargam=0.5$ as
in~\cite{Biktasheva-etal-2009}. 

The \textcite{Barkley-1991} kinetics is given by
\begin{align}
  f(u,v) &= \parc^{-1} u \, (1-u)(1-(v+\parb)/\para),       \nonumber\\
  g(u,v) &= u-v,                                            \nonumber
\end{align}
with parameter values $\para=0.7$, $\parb=0.01$ and $\parc=0.025$, as
in~\cite{Henry-Hakim-2002}. 
Note that both $\paralp$ and
$\parc$ are called $\epsilon$ in \cite{Winfree-1991} and
\cite{Barkley-1991} respectively; however we use $\epsilon$ for the
small parameter in the perturbation theory.

\subsection{Response functions computations}

\dblfigure{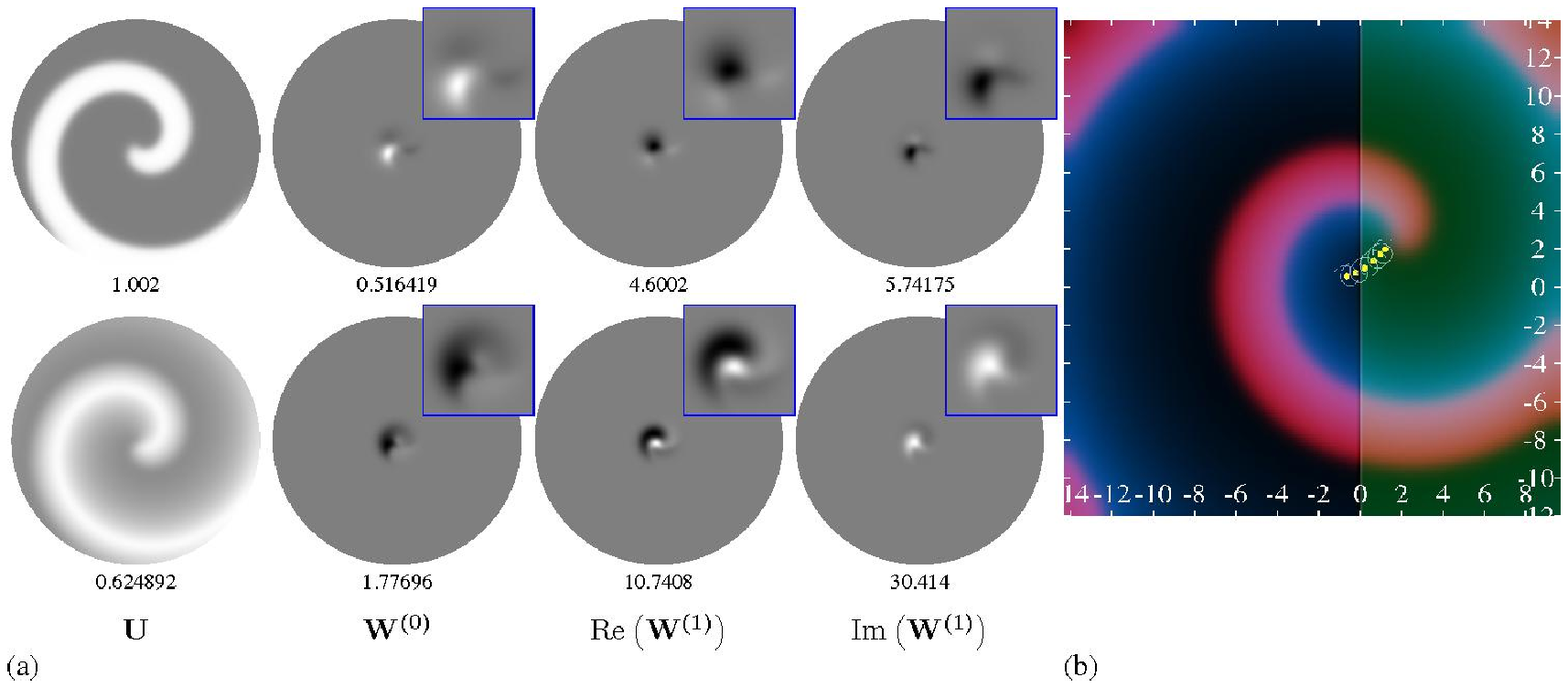}{(color online) %
  (a) 
  Solutions of the nonlinear problem \eq{SW-own} and the adjoint linearized problem
  (\ref{mus},\ref{Lp}),
  \ie\ the response functions, as density plots. Barkley model,
  $\rp=12.8$, $\Nr=1280$. 
  Numbers under the density plots are their amplitudes $\Ampl$: 
  white of the plot corresponds to the value $\Ampl$ and black corresponds
  to the value $-\Ampl$ of the designated field.
  Upper row: 1st components ($u$), lower row: 2nd components ($v$).
  The central areas of $\RF{n}$, $n=0,1$, are also shown
    magnified in the small corner panels.
  (b) Snapshot of spiral wave in the Barkley model ($u$: red colour
  component, $v$: blue colour component),
  drifting in a stepwise inhomogeneity of paramer $\parc$ (green
  colour component). The thin white line is the trace of the tip of
  the spiral in the course of a few preceding rotations. Yellow
  circles are positions of the centres calculated as period-averaged
  positions of the tip. 
}{rfs}

For both the FitzHugh-Nagumo and the Barkley models, the response
functions and the Goldstone modes have been computed using the methods
described in \cite{Biktasheva-etal-2009}. The discretization is on
disks of radii from $\rp=12$ up to $\rp=50$, using $\Nt=64$ of
discretization intervals in the angular direction and a varying number
$\Nr$ of discretization intervals in the radial direction, up to
$\Nr=1280$. The components of the spiral wave solution and its
response functions for Barkley model are shown in \fig{rfs}(a).
Similar pictures for the FitzHugh-Nagumo model can be found
in~\cite{Biktasheva-etal-2009}.

\subsection{Perturbations}

We considered similar types of perturbations $\epsilon\h(\u,\r,t)$ in
both FitzHugh-Nagumo and Barkley models, both for theoretical
predictions based on response functions and in numerical simulations.
Specifically, the perturbations were taken to have the following forms.

\paragraph{Resonant drift}
\begin{equation}
\h(\u,\r,t) = \Mx{1\\0} \cos(\omega t),     \label{hres-method}
\end{equation}
where $\omega$ is the
angular velocity of the unperturbed spiral obtained as part of the spiral wave solution for the equation \eq{SW-own}.

\paragraph{Electrophoretic drift}
\begin{equation}
\h(\u,\r,t) = \Mx{1&0\\0&0} \df{\u}{x} . \label{helec-method}
\end{equation}

\paragraph{Spatial parametric inhomogeneities.}

As set out in \secn{theory}, a spatial dependence of a parameter
$\param$ of the kinetic terms in the form
$\param(\r)=\param_0+\param_1(\r)$, $|\param_1|\ll|\param_0|$
corresponds to the perturbation
\begin{equation}
  \h(\u,\r,t) = \dfdp(\u,\param_0) \, \param_1(\r).         \label{spat-param}
\end{equation}
For each of the two models, we consider inhomogeneities in all three
parameters, namely $\param\in\{\paralp,\parbet,\pargam\}$ for the
FitzHugh-Nagumo model, and $\param\in\{\para,\parb,\parc\}$ for the
Barkley model. The ``linear gradient'' inhomogeneity is of the form
\begin{equation}
  \param_1 = x-x_0,
\end{equation}
where $x_0$ is chosen to be in the middle of the computation box and close
to the initial centre rotation of the spiral wave.

The ``stepwise'' inhomogeneity is of the form
\begin{equation}
  \param_1 = \Heav(x-\xs)-\frac12 ,                         \label{stepwise}
\end{equation}
where $\xs$ is varied and chosen with respect to the initial centre
of rotation of the spiral wave. The $-\frac12$ term is added to make
the perturbation symmetric (odd) about $x=\xs$, to minimize the
inhomogeneity impact on the spiral properties while near the step. As
it can be easily seen, within the asymptotic theory, this term only
affects the frequency of the spiral but not its spatial drift.

The ``disk-shape'' inhomogeneity is of the form
\begin{equation}
  \param_1 = \Heav(\Ri-|\r-\rd|),                           \label{disk-shape}
\end{equation}
where the position of the centre of the disk $\rd=(\xd,\yd)\T$ is varied
and chosen with respect to the initial centre of rotation of the
spiral wave.

\subsection{Drift simulations}

Simulations have been performed using forward Euler timestepping on uniform
Cartesian grids on square domains with non-flux boundary conditions
and five-point approximation of the Lapacian. The space discretization
step $\dx$ has been varied between $\dx=0.03$ and $\dx=0.1$, and time
discretiation step $\dt$ maintained as $\dt=\frac15\dx^2$. The
tip of the spiral is defined as the
intersections of isolines $u(x,y)=u_*$ and $v(x,y)=v_*$, 
and the angle of $\nabla u$ at the tip
  with respect to $x$ axis is taken as its orientation.
We use
$(u_*,v_*)=(0,0)$ for the FitzHugh-Nagumo model and
$(u_*,v_*)=(1/2,\para/2-\parb)$ for the Barkley model.

\subsection{Processing the results}

For coarse comparison, we use the trajectories of the instantanous
rotation centre of the spiral wave. They are directly predicted by the
theory. In simulations, they are calculated by averaging the position
of the tip during full rotation periods, defined as the intervals when
the orientation makes the full circle $(-\pi,\pi]$, see \fig{rfs}(b).

For finer comparison, we fit the raw tip trajectories, \ie\ we use
theoretical predictions including the rotation of the spiral. That is,
if the theory predicts a trajectory of the centre as
$R=R(t;A,B,\dots)\in\Complex$ (a circle for resonant drift and a
straight line for electrophoretic or linear gradient inhomogeneity
drifts) depending on parameters $A,B,\dots$ to be identified, then the
trajectory of the tip is assumed in the form $\Rtip(t)=R(t;A,B,\dots)
+ \Rcore e^{i(\omega t+\iniphase)}$ where $\Rcore\in\Real$ is the tip
rotation radius, $\omega\in\Real$ is the spiral rotation frequency and
$\iniphase\in\Real$ is the initial phase. The parameters $\Rcore$,
$\omega$ and $\iniphase$ are added to the list $A,B,\dots$ of the
fitting parameters.

\section{Results}
\seclabel{results}

\subsection{Simple drifts}

\dblfigure{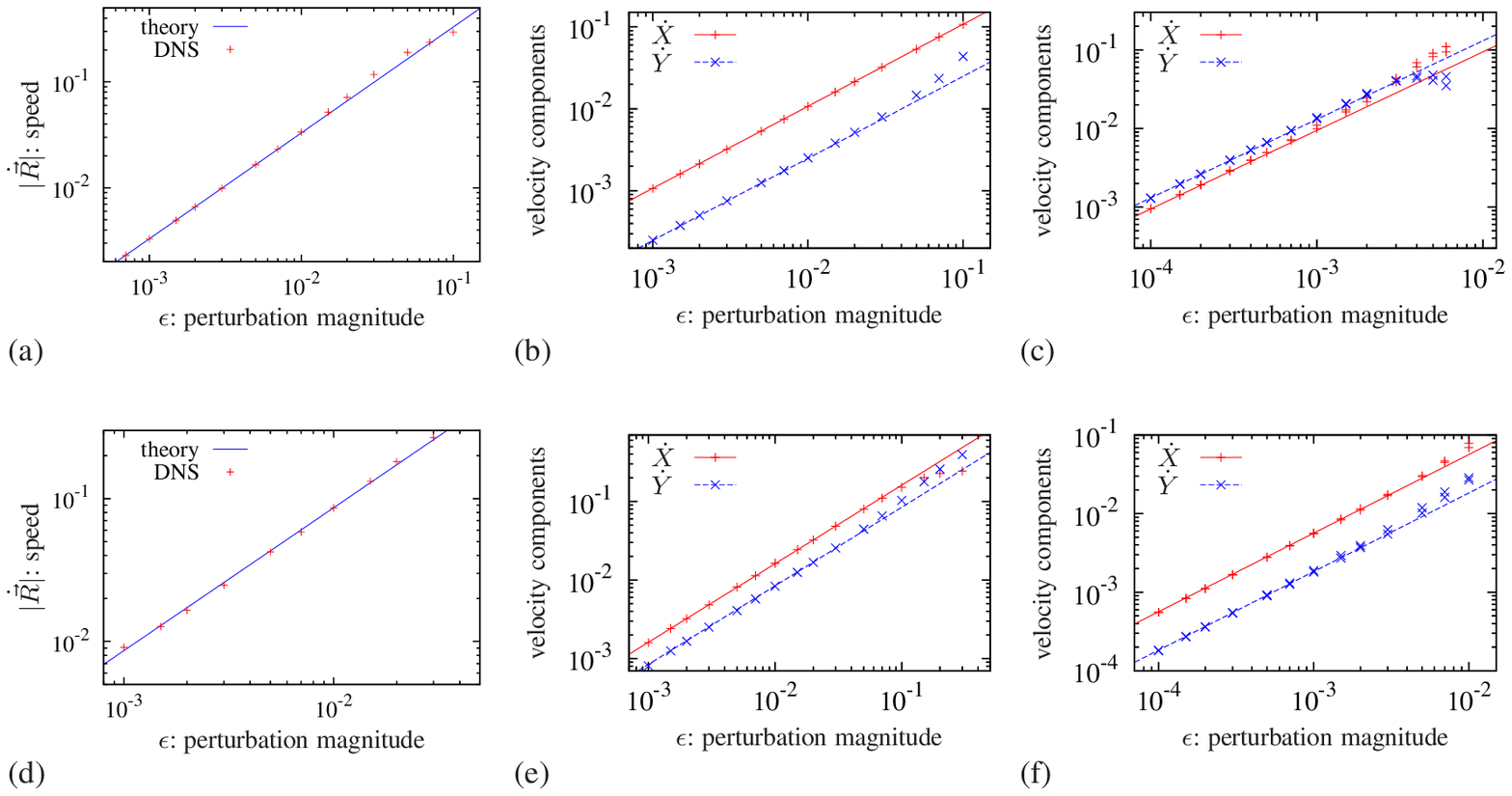}{(color online) %
  Drift speeds as functions of corresponding perturbation amplitudes. 
  Top row: FitzHugh-Nagumo model. Bottom row: Barkley model. 
  First column: resonant drift. Second column: electrophoretic drift. 
  Third column: drift in linear gradient inhomogeneity, namely 
  (c) with respect to parameter $\parbet$ in the FHN model, and
  (f) with respect to parameter $\para$ in the Barkley model. 
  In the second and the third columns, the symbols represent
  simulations and the lines represent theoretical predictions.
  Numerical parameters: 
  (a) $\dx=\dr=0.1$, $\rp=50$,
  (b) $\dx=\dr=0.1$, $\rp=25$,
  (c) $\dx=0.08$, $\dr=0.02$, $\rp=25$, 
  (d) $\dx=\dr=0.05$, $\rp=25$,
  (e) $\dx=\dr=0.02$, $\rp=12.5$,
  (f) $\dx=\dr=0.06$, $\rp=24$.
}{speeds}

\Fig{speeds} shows a comparison between the theoretical predictions for the
simple drifts and the results of direct numerical simulations of
various perturbation amplitudes $\epsilon$. The simple drifts include
the resonant drift, the electrophoretic drift and the drift in the
linear parametric gradient with respect to one arbitrarily selected
parameter. 

For the resonant drift, the motion equations given by \eq{res-speed},
\eq{res-direction} and \eq{res-direction-2} can be summarized, in
terms of complex coordinate $R=X+\i Y$, as
\begin{equation}
  \Df{R}{t} = e^{\i\Phi}  \resdrspeed, \qquad
  \Df{\Phi}{t} = \resdrangvel,
\end{equation}
where
$\resdrspeed=\frac{1}{2}\left|\epsilon\inner{\RF{1}}{\Ampres}\right|$
is predicted by the theory at leading order, and
$\resdrangvel=\O(\epsilon^2)$ is not, and we only know its expected
asymptotic order. The theoretical trajectory is a circle of radius
$\resdrspeed/\resdrangvel$, and the spiral drifts along it with the
speed $\resdrspeed$. In the simulations, we determined both the radius
and the speed by fitting. The speed is used for comparison and the
radius is ignored.

For the other two types, electrophoretic drift and linear gradient
inhomogeneity drift, the theory predicts drift at a straight line,
according to \eq{forces_electr} and \eq{speed_gradient}
respectively. In these cases, we measure and compare the $x$ and $y$
components of the drift velocities separately.

For numerical comparison in the case of linear gradient inhomogeneity,
we chose a pieces of trajectories not too far from $x=x_0$, selected
empirically to achieve a satisfactory quality of fitting.  

A common feature of all graphs is that at small enough
$\epsilon$, there is a good agreement between theory and
simulations.
As expected, differences appears for larger $\epsilon$ with the  
disagreement occurring sooner (for smaller values of drift
speed) for the linear gradient inhomogeneity drift. This is related to
an extra factor specific to the inhomogeneity-induced drift: the
properties of the medium where they matter, \ie\ around the core of
the spiral, changes as the spiral drifts. Since we require a certain
number of full rotations of the spiral for fitting, faster drift meant
longer displacement along the $x$ axis and more significant change of
the spiral properties along that way, which in turn affects the
accuracy of the fitting.

\subsection{Numerical convergence}

\dblfigure{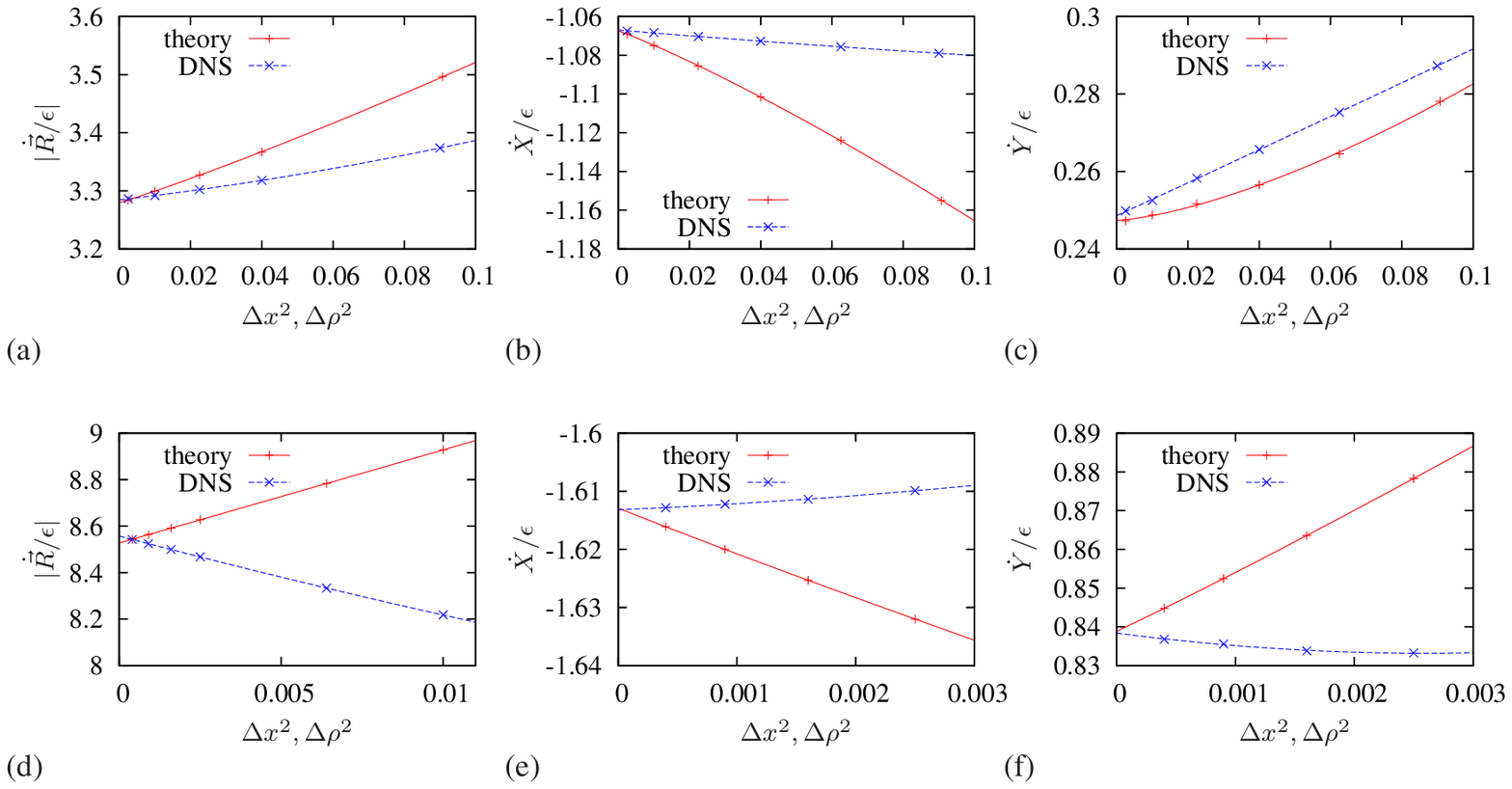}{(color online) %
  Numerical convergence of drift forces.
  Top row: FHN. Bottom row: Barkley.
  First column: resonant drift. 
  Second and third columns: $x$ and $y$ components of the
  electrophoretic drift.
  The forces were measured for 
  (a) $\epsilon=10^{-3}$,
  (b,c) $\epsilon=10^{-2}$,
  (d) $\epsilon=5\cdot10^{-3}$,
  (e,f) $\epsilon=10^{-2}$.
}{conv}

\Fig{conv} illustrates numerical convergence of results with
discretization parameters. We consider the simple drift cases and
focus on forces, defined as the drift speed/velocity per unit
perturbation amplitude $\epsilon$. 
The discretization parameter that primarily dictates the accuracy of 
solutions is a spatial discretization step: $\dx$ in the simulations
and the radial discretization step $\dr$ in the response functions
calculations.

In simulations, the forces are determined for values
of $\epsilon$ well within the linear range as determined in
\Fig{speeds}. These are calculated for different values of the space
discretization step $\dx$, where the time discretization changed
simultaneously so that the ratio $\dt/(\dx)^2$ remaines constant.

In theoretical predictions, the forces are given by the values of
the corresponding integrals of response functions as described by
\secn{theory}, and we have calculated the response functions and the
corresponding integrals with various values of the radius discretization
steps $\dr$.

Our discretization in both the theoretical and stimulations cases is
second order in $\dx$ and in $\dr$, so one would expect to see linear
dependence of the drift forces on the squares of these
discretization steps, $(\dx)^2$ and $(\dr)^2$, at least for the values of these
steps small enough. This is indeed what is observed.

We have gone further and extrapolated the calculated theoretical and simulation
values of forces to zero $\dr$ and $\dx$ respectively, based on the
expected numerical convergence properties. Such extrapolation gives the
values of the forces which differ from the exact value only due to
other, smaller discretization errors, which are: angular discretization and
restriction to the finite domain in the theoretical predictions, and
second-order corrections in $\epsilon$ and the boundary effects in the
simulations. Comparison of such extrapolated data shows a very good
agreeement between theory and direct numerical simulations (DNS) which is illustrated in
Table~\ref{tab:fitting}.
Note that the values for \fig{conv}(e) and \fig{conv}(f) are also in
good agreement with the results of \cite{Henry-Hakim-2002}. 

\begin{table*}[htbp]
  \begin{tabular}{|l|l|l|l|} \hline
    Graph & Theory & DNS & Discrepancy \\\hline
    \fig{conv}(a) & $ 3.2795 + 1.8183 \dr^2 + 1.8928 \dr^3 $ 
                  & $ 3.2844 + 0.7166 \dx^2 + 3.0663 \dx^4$ & 0.15\% \\\hline
    \fig{conv}(b) & $ -1.0673 - 0.6418 \dr^2 - 1.0733 \dr^3 $ 
                  & $ -1.0670 - 0.1513 \dx^2 + 0.2053 \dx^4$ & 0.03\% \\\hline
    \fig{conv}(c) & $ 0.2474 + 0.0141 \dr^2 + 1.0697 \dr^3 $ 
                  & $ 0.2486 + 0.4224 \dx^2 + 0.0789 \dx^4$ & 0.49\% \\\hline
    \fig{conv}(d) & $ 8.5277 + 39.5595 \dr^2 + 4.9401 \dr^3 $ 
                  & $ 8.5574 - 36.7653 \dx^2 + 278.0169 \dx^4$ & 0.35\% \\\hline
    \fig{conv}(e) & $ -1.6129 - 8.2600 \dr^2 + 12.3530 \dr^3 $ 
                  & $ -1.6132 + 0.8648 \dx^2 + 178.7182 \dx^4$ & 0.02\% \\\hline
    \fig{conv}(f) & $ 0.8389 + 14.1446 \dr^2 + 32.6213 \dr^3 $ 
                  & $ 0.8384 - 3.9945 \dx^2 + 769.7281 \dx^4$ & 0.06\% \\\hline
  \end{tabular}
  \caption{Fitting of numerical convergence of theoretical and simulation data}
  \label{tab:fitting}
\end{table*}

For the extrapolation, we fitted the numerical data with the expected
numerical convergence dependencies, which were different for theoretical
calculations and for the simulations. In simulations, the central
difference approximation of the Laplacian means that the next term after
$(\dx)^2$ is $(\dx)^4$. The expected error due to time derivative
discretization is a power series in $\dt\sim(\dx)^2$, hence the next term
there after $(\dx)^2$ is again $(\dx)^4$. The situation is different in the
response functions calculations as there is no symmetry in the
approximation of $\rho$-derivatives, therefore we expect that in the
theoretical convergence, the next term after $(\dr)^2$ is $(\dr)^3$. We note,
however, that approximation of both theoretical and simulation data with
similar dependencies, be it with a cubic or a quartic third term, gave very
similar results.

\subsection{Drift near stepwise inhomogeneity}

\dblfigure{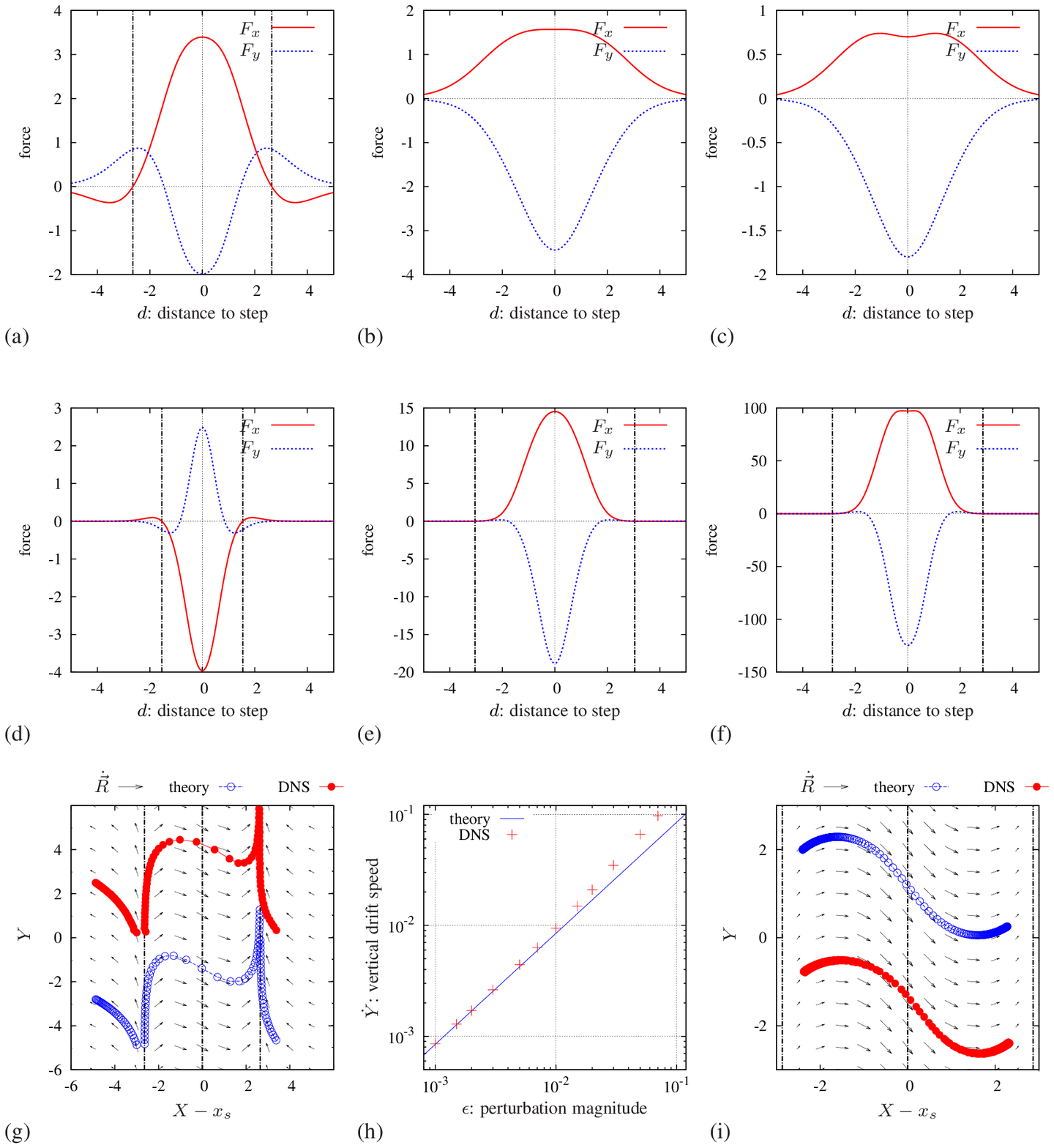}{(color online) %
  Drift in stepwise inhomogeneity (\ref{spat-param},\ref{stepwise}). 
  First row: theoretical predictions for the drift forces components as
  functions of the distance to the steps,
  $d=X-\xs$,
  in parameters
  (a) $\alpha$, 
  (b) $\beta$ and 
  (c) $\gamma$, in the FitzHugh-Nagumo model. 
  Second row: same for Barkley model, steps in parameters 
  (d) $\para$, 
  (e) $\parb$ and
  (f) $\parc$. 
  Third row: comparison of theoretical predictions with DNS.
  (g) A phase portrait of the drift in the FitzHugh-Nagumo model,
  in theory, \eq{step-velocity}, and DNS,
    (\ref{RDS_pert},\ref{spat-param},\ref{stepwise}),
  with
  a step inhomogeneity of parameter $\alpha$ (corresponds to panel
  (a)), at $\epsilon=10^{-2}$. Shown are the theoretical vector field
  (black arrows; the lengths are nonlinearly scaled for
  visualization), a selection of theoretical trajectories (red filled
  circles) and a selection of numerical trajectories (blue open
  circles) of the centres of the spiral waves. Trajectories are
  arbitrarily shifted in the vertical direction for visual
  convenience.  Dashed-dotted vertical lines correspond to the root of
  the theoretical horizontal component of the speed, and the location
  of the step $X-\xs=0$.
  (h) Speed of the established vertical movement along the stepwise
  inhomogeneity as in panel (g), as a function of inhomogeneity
  strength. 
  (i) A phase portrait of the drift in the Barkley model
  with a step inhomogeneity of parameter $\parc$ %
  (corresponds to panel (f)), at $\epsilon=3\cdot10^{-4}$. %
  Notation is the same  as in panel (g). 
}{step}

The theoretical predictions for stepwise inhomogeneity, 
(\ref{spat-param},\ref{stepwise})
and disk-shaped
inhomogeneity considered next, are more complicated than the simple forms of
drift considered up to this point. Because now the medium is inhomogeneous in
the presence of the perturbation, the velocity depends on the instant position
of the spiral center and as a result the spiral trajectories can be quite
complex. 
Qualitative comparisons between theory and simulations can be made in the
general case, but for detailed quantitative comparison we focus on the cases
where the theory predicts simple attractors, \eg ~a straightforward drift along the
step for the stepwise inhomogeneity.

Equations \eq{step-velocity} give a system of two first-order autonomous 
differential equations for $X=\Re{R}$ and $Y=\Im{R}$,
\begin{eqnarray}
  \dtime{X} &=& \epsilon \Fx(X-\xs), \nonumber \\
  \dtime{Y} &=& \epsilon \Fy(X-\xs), \label{stepODE}
\end{eqnarray}
where
\begin{align}
  \Fx(\dists) &= 
  \frac{1}{\pi} \;
  \int\limits_0^{2\pi} \int\limits_{\left|\dists\right|}^\infty \; 
  \Re{ \w{1}(\rho,\theta) \e^{-\i\theta}} \; 
  \sqrt{\rho^2-\dists^2} \; 
  \,\d\rho\,\d\theta , \label{Fx} \\
  \Fy(\dists) &= 
  \frac{1}{\pi} \;
  \int\limits_0^{2\pi} \int\limits_{\left|\dists\right|}^\infty \; 
  \Im{ \w{1}(\rho,\theta) \e^{-\i\theta}} \; 
  \sqrt{\rho^2-\dists^2}
   \,\d\rho\,\d\theta . \label{Fy}
\end{align}
The right-hand sides of system \eq{stepODE} depend only on $X$ but not
on $Y$, that is, the system is symmetric with respect to translations
along the $Y$ axis. For this reason, the roots
$\dists_*:\Fx(\dists_*)=0$ provide invariant straight lines along the
$Y$-axis. An invariant line $\{(\xs+\dists_*,Y)|Y\in\Real\}$ will be
stable if $\epsilon\Fx'(\dists_*)<0$ and unstable if
$\epsilon\Fx'(\dists_*)>0$. Note that the stability of invariant lines
reverses with a change of sign of $\epsilon$ and also that
$\Fx(\dists)$ is an even function. Hence if $\epsilon\ne0$ and
$\dists_*\ne0$ then either $\{\xs+\dists_*,Y\}$ or $\{\xs-\dists_*,Y\}$
will be an attracting invariant set.

\Fig{step}(a--f) show the theoretical predictions for the drift
forces, \ie\ velocity components per unit perturbation magnitude
$\epsilon$, $\Fx(\dists)$ and $\Fy(\dists)$, on the distance
$\dists=X-\xs$ from the instant spiral centre to the step. This is
done for both FitzHugh-Nagumo and Barkley models, for steps in each of
the three parameters in these models. The roots of $\Fx(\dists)$ are
specially indicated. One can see from the given six examples, that
existence of roots of $\Fx()$ is quite a typical, albeit not a
universal, event.

The qualitative predictions of the theory about a stable invariant
line are illustrated by \fig{step}(g) where we present results of
numerical integration of the ODE system~\eq{stepODE} and the results
of direct numerical simulation of the full system. In the example
shown, the positive root $\dists_*\approx2.644$ of $\Fx$ is stable and
the negative root $-\dists_*$ is unstable. Hence the theoretical
prediction for different initial conditions are: %
\begin{description} %
\item{for $X(0)>\xs-\dists_*$ and not too big}, the spiral wave will approach
  the line $X=\xs+\dists_*$ and drift vertically along it with the
  speed $\epsilon\Fy(\dists_*)\approx0.8468\epsilon$; %
\item{for $X(0)<\dists_*$}, the spiral wave will drift to the left with ever
  decreasing speed, until its drift is no longer detectable; %
\item{for big $|X(0)|$}, the drift will not be
  detectable from the outset. %
\end{description}

As seen in \fig{step}(g) this is indeed what is observed, both for the
theoretical and for the DNS trajectories, and the visual similarity
between theoretical and DNS trajectories is an illustration of the
validity of the qualitative predictions of the theory.

Since the generic drift is non-stationary, a quantitative comparison
for typical trajectories is difficult. However, the drift along the
stable manifold $X=\xs+\dists_*$ is stationary with vertical velocity
given by $\epsilon\Fy(\dists_*)$ so a comparison is easily made using
the same methods
as in the case of ``simple'' drifts considered in the previous
subsections. The results are illustrated in \fig{step}(h). As
expected, we see good agreement between the theory and the DNS for
small $\epsilon$.

The phenomenological predictions are different for the case when
$\Fx(\dists)$ has no roots, or when its roots are so large that
$|\Fy(\dists_*)|$ is so small that the drift cannot be detected in
simulations. In such cases, the theoretical predictions for different initial conditions are:
\begin{description} %
\item{for $|X(0)|$ not too large}, the spiral wave will move with varying
  vertical velocity component but always in the same
  horizontal direction (to the right if $\epsilon\Fx(0)>0$),
  eventually with ever decreasing speed, until its drift is no longer
  detectable; %
\item{for $|X(0)|$ too large}, the drift will not be
  detectable from the outset. %
\end{description}
This prediction is confirmed by simulations, as illustrated in
\fig{step}(i), where we have chosen the case of inhomogeneity in
parameter $\parc$ in Barkley model, for which the smallest positive
root is $\dists_*\approx2.867$, which gives
$\Fy(\dists_*)\approx0.1632$. This value should be compared to
$\Fx(0)\approx48.42$ and $\Fy(0)\approx119.4$. Note also that to get
the drift velocities, $\Fx$ and $\Fy$ should be multiplied by
$\epsilon$ which should be much smaller than $\parc_0=0.025$. So when
the spiral is further than $|X-\xs|\sim2$ from the step, the drift is
very slow and hardly noticeable, even though according to the theory,
there should be stable vertical drift around $X=\xs+\dists_*$, which is
too slow to be observed in normal simulations.

\subsection{Drift near disk-shape localized inhomogeneity}

\dblfigure{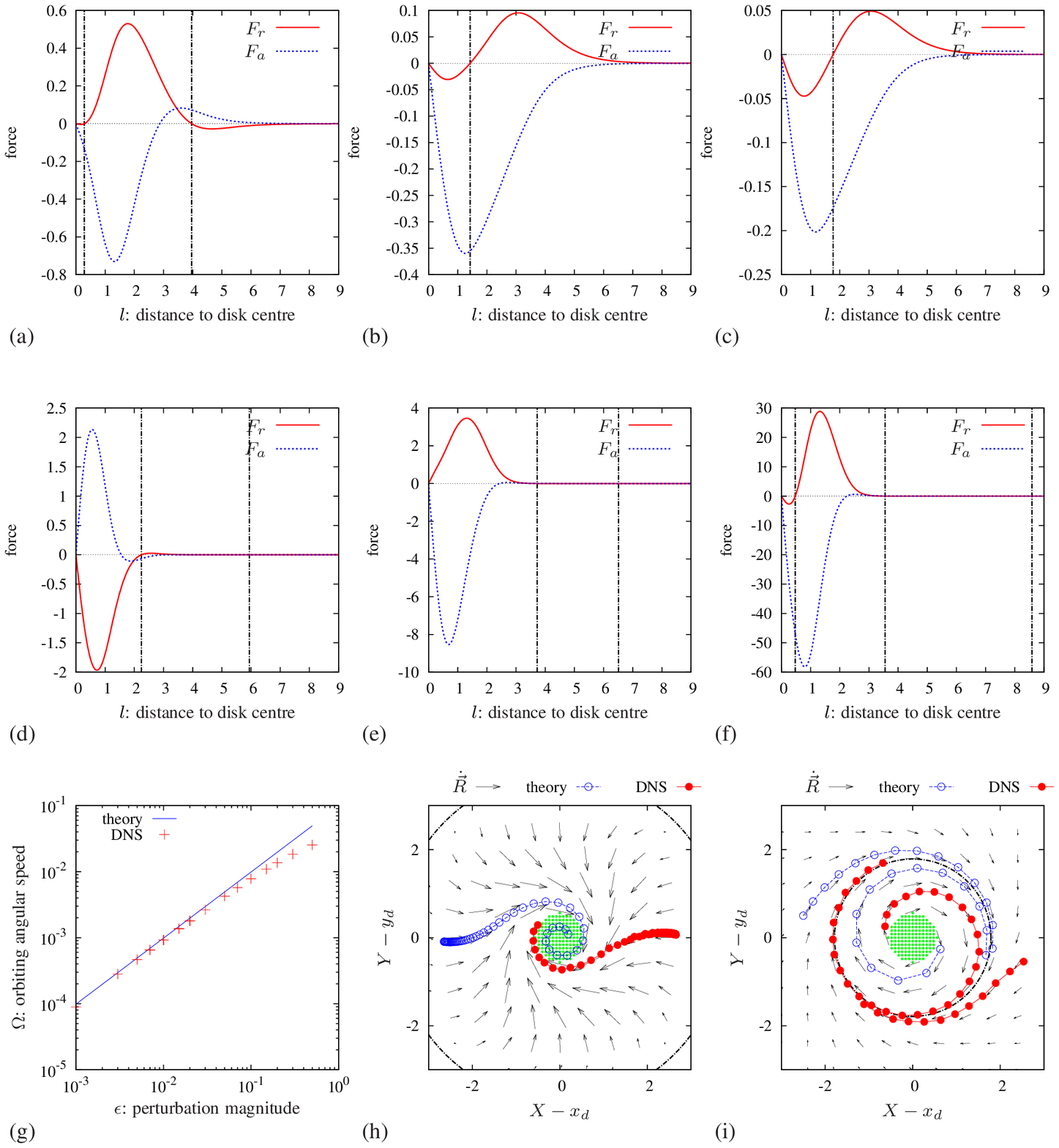}{(color online) %
  Drift around disk-shape inhomogeneity (\ref{spat-param},\ref{disk-shape})
  of radius $\Ri=0.56$. 
  First row: theoretical predictions for the drift speed components as
  functions of the distance to the disk centre, 
  $l=\left((X-\xd)^2+(Y-\yd)^2\right)^{1/2}$,
  for inhomoegenity in parameters
  (a) $\alpha$, 
  (b) $\beta$ and 
  (c) $\gamma$, in the FitzHugh-Nagumo model. 
  Second row: same for Barkley model, steps in parameters 
  (d) $\para$, 
  (e) $\parb$ and
  (f) $\parc$. 
  Third row: comparison of theoretical predictions with DNS.
  (g) Angular speed of the established orbital movement around the 
  inhomogeneity site as on panel (i), as a function of inhomogeneity strength. 
  (h) A phase portrait of the drift in the Barkley model 
  in theory, \eq{disk-velocity}, and DNS,
    (\ref{RDS_pert},\ref{spat-param},\ref{disk-shape}),
  with
  disk-shape inhomogeneity (green) of parameter $\parb$ (corresponds
  to panel (e)), at $\epsilon=10^{-2}$. Shown are the theoretical vector field (black arrows;
  the lengths are nonlinearly scaled for visualization), a selection
  of theoretical trajectories (red filled circles) and a selection of
  numerical trajectories (blue open circles) of the centres of the
  spiral wave. Dash-dotted circles correspond to the roots of the theoretical
  radial component of the drift force.
  (i) A phase portrait of the drift in the FitzHugh-Nagumo model with
  inhomogeneity of parameter $\gamma$ (corresponds to panel (c)),
  $\epsilon=0.3$. Notation is the same as in panel (h).
}{disk}

The theoretical predictions for the disk-shaped inhomogeneity
(\ref{spat-param},\ref{disk-shape}),
are more complicated but also more interesting. 
The theoretical spiral motion equation
\eq{disk-velocity} has a rotational, rather than the translational symmetry
of the stepwise inhomogeneity. In polar coordinates
$(\distd,\angled)$ centered at the center of the inhomogeneity, so that
$R=\xd+\i\,\yd+\distd\e^{\i\angled}$, equation \eq{disk-velocity} can
be rewritten in the form
\begin{align}
  \dtime{\distd} & = - \epsilon \Fr(\distd), \nonumber\\
  \distd \dtime{\angled} &= \epsilon \Fa(\distd), \label{diskODE}
\end{align}
where $\Fr$ and $\Fa$ are the radial and azimuthal components of the drift
force, given by
\begin{align}
  \Fr = \int\limits_{0}^{2\pi} \int\limits_{\left| \distd-\Ri \right|}^{\distd + \Ri} & 
    \Re{ \w{1}(\rho,\theta) \e^{-i\theta}}
\nonumber\\&
    \sqrt{1 - \left(\frac{\rho^2 + \distd^2 - \Ri^2}{2\distd\rho} \right)^2} \;
     \,\rho\,\d\rho\,\d\theta , \label{Fr}\\
  \Fa = \int\limits_{0}^{2\pi} \int\limits_{\left| \distd-\Ri \right|}^{\distd + \Ri} &
    \Im{ \w{1}(\rho,\theta) \e^{-i\theta}}
\nonumber\\&
    \sqrt{1 - \left(\frac{\rho^2 + \distd^2 - \Ri^2}{2\distd\rho} \right)^2} \;
     \,\rho\,\d\rho\,\d\theta.   \label{Fa}\\
\end{align}
The minus sign in the first equation of \eq{diskODE} comes from the
fact that in \eq{disk-velocity}, the origin was placed at the instant
rotation centre of the spiral, and the position of inhomogeneity is
determined with respect to it, where as now we do the other way round:
the origin is at the centre of inhomogeneity and the current position
of the spiral rotation centre is determined with respect to it.

In the system \eq{diskODE}, the axial symmetry is manifested by the
fact that the right-hand sides of \eq{diskODE} depend on $\distd$ but
not on $\angled$, and the equation for $\distd$ is a closed one. Hence
roots $\distd_*$ of $\Fr(\distd_*)=0$ represent invariant sets, which
in this case are circular orbits. The movement along those orbits will
have a linear speed $\epsilon\Fa(\distd_*)$ and angular speed
$\Omega=\epsilon\Fa(\distd_*)/\distd_*$. The stability of these orbits is
determined by the sign of $\epsilon\Fr'(\distd_*)$: stable for
positive and unstable for negative. Unlike the case of the stepwise
inhomogeneity, now we do not have any mirror symmetries, as only
positive $\distd$ make sense, therefore for a given root $\distd_*$ a
stable circular orbit is guaranteed only for one sign of $\epsilon$
but not the other.

\Fig{step}(a--f) show the theoretical predictions for the drift
forces
$\Fr(\distd)$ and $\Fa(\distd)$. This is done for both FitzHugh-Nagumo
and Barkley models, for inhomogeneities in each of the three
parameters in these models. The roots of $\Fr(\distd)$ are specially
indicated. One can see from the six given examples that existence of
roots of $\Fr()$ is rather common and often
there is more than one root, $\distd_j$, such that
$0=\distd_0<\distd_1<\distd_2<\dots$.

The qualitative predictions of the theory are illustrated in
\fig{disk}(h,i) where we present results of numerical integration of
the ODE system~\eq{diskODE} and the results of direct numerical
simulations of the full system. In the example shown in \fig{disk}(h),
the predictions are given by \fig{disk}(e), which say that the
smallest orbit has radius $\distd_1\approx3.724$, with the orbital
speed $\Fa(\distd_1)\approx0.003938$, which is rather small compared
to $\max(|\Fr(\distd)|\approx3.458$ and
$\max(|\Fa(\distd)|\approx8.534$ and hardly observable in numerical
simulations. Hence in this case, the radial component of the drift
speed $\Fr(\distd)$ is effectively constant-sign, and for negative
$\epsilon$, one should observe repulsion of the spiral wave from the
inhomogeneity until it is sufficiently far from it, $\distd\sim3$, to
stop feeling it, and for positive $\epsilon$, the spiral wave will be
attracted towards the centre of inhomogeneity from any initial position
$\distd\simeq3$. This is indeed what is observed in simulations shown
in \fig{disk}(h) where the case of $\epsilon>0$ is shown, and the
centre of the inhomogeneity, $\distd=\distd_0$, is attracting for the
spiral wave.

In the example shown in \fig{disk}(i), the inhomogeneity centre
$\distd_0=0$ is repelling. Instead, the first orbit of radius
$\distd_1\approx1.7722$ is attracting. The perturbation amplitude
$\epsilon=0.3$ in this case is quite large and comparable with the value
$\gamma_0=0.5$ of the perturbed parameter itself. We see that although
the numerical correspondence between theory and DNS in this case is
not very good (note the distances between the open circles and between
the filled circles), the qualitative prediction of orbital movement
remains impeccable. As expected, the numerical correspondence becomes
good for smaller values of $\epsilon$, see \fig{disk}(g).

\section{Discussion}
\seclabel{discussion}
\setcounter{paragraph}{0}

We have considered symmetry breaking perturbations of three different kinds:
time-translation symmetry breaking, that is homogeneous in space and periodic
in time (``resonant drift''); rotational symmetry breaking through differential
advective terms (``electrophoretic drift''); and spatial translation symmetry
breaking through space-dependent inhomogeneities (``inhomogeneity induced
drift'').  The latter type includes three sub-cases cases: a linear parametric
gradient, a stepwise parameter between to half plane, and a parameter
inhomogeneity localized within a disk.

\paragraph{Quantitative: drift velocity.} 
We have demonstrated that asymptotic theory gives accurate
predictions for spiral drift: in some cases the discrepancy between the
theory and the direct simulations was as low as 0.02\%. The
discrepancy is affected by the numerical discretization parameters, both for
the direct simulations and for response function computations, and by
the magnitude of the perturbation.

\paragraph{Qualitative: attachment and orbiting.}
In the more complicated cases of spatial inhomogeneity, the response functions 
allow us to predict qualitatively different regimes of spiral motion, which we  have been able to confirm by direct simulations.

In the presence of a stepwise inhomogeneity, the centre of spiral wave 
rotation may
either be attracted to one side of the step where it gradually
``freezes'', or it may get attached to the step and drift along it
with the constant velocity. In the latter case, the speed of the drift
is proportional to the inhomogeneity strength, whereas the distance at
which the attachment happens, does not not depend on  the inhomogeneity
strength at leading
order. If the sign of inhomogeneity is inverted, the attachment occurs
on the opposite side of the step and proceeds in the opposite direction.

In disk-shape inhomogeneity, the situation is somewhat similar but
more interesting. The spiral wave may be attracted towards the centre
of the disk, or repelled from it. It may also be attracted to or
repelled from one or more circular orbits. The drift velocity 
along the orbits is proportional to the strength of the inhomogeneity,
whereas the radii of orbits do not depend on it at leading
oder. The repulsion changes to attraction and vice versa, with the
change of the sign of the inhomogeneity.

\paragraph{Prevalence of attachment and orbiting}
The possibilities of attachment to the step inhomogeneity and orbital
movement for the disk-shape inhomogeneity are both related to the change
of sign of the integrals of the translational response functions, which in
turn are possible due to changes of sign of the components of those
response functions. Not suprisingly, there is a certain correlation
between these phenomena. The graphs \fig{disk}(a--f) may be viewed as
deformed versions of the corresponding graphs
\fig{step}(a--f). Respectively, positive roots of $\Fx(\dists)$ in
\fig{step}(a,d,e,f) have corresponding roots of $\Fr(\distd)$ in
\fig{disk}(a,d,e,f). However, the integrals in equations \eq{Fx} and
\eq{Fr} are only similar but not identical, and the above
correspondence between the roots is not absolute: the roots of
$\Fr(\distd)$ in \fig{disk}(b,c) and the smaller roots of this
function in \fig{disk}(a,f) have no correspondences in \fig{step}.
Overall, based on results considered, orbital motion around a
localized inhomogeneity seems to be more prevalent than attachment to
a stepwise inhomogeneity. Moreover, the typical situation seems to be
that there are multiple stationary orbits around a disk
inhomogeneity. We have already discussed this situation in our recent
preliminary short communication~\cite{OMS} where we have also
illustrated how for the initial conditions between two stationary
orbits, the spiral wave launched into one orbit or the other depending
on the sign of the inhomogeneity.

The possibility of orbital drift, related to a change of sign of an
equivalent to the function $\Fr(\distd)$, has been discussed at a speculative
level in~\cite{LeBlanc-Wulff-2000}. The sign change of translational
response functions was observed in oscillatory media described by
CGLE~\cite{Biktasheva-PhD-Russ, Biktasheva-Biktashev-2001}. The
examples we consider here suggest that this theoretical possibility
is in fact quite often realized in excitable media, and even multiple
orbits are quite typical. Theoretical reasons for this
prevalence are not clear at present.  As stated in~\cite{OMS}, 
the prevalence of multiple orbits may be understood in
terms of asymptotic theories involving further small parameters. So,
the version of kinematic theory of spiral waves suggested
in~\cite{Elkin-Biktashev-1999} produces an equivalent of response
functions, which is not only quickly decaying, but also periodically
changing sign at large radii, with an asymptotic period equal to the
quarter of the asymptotic wavelength of the spiral wave. Other
variants of the kinematic theory,
\eg~\cite{Mikhailov-etal-1994,Hakim-Karma-1999} did not, to our
knowledge, reveal any such features on a theoretical level. However,
numerical simulations of kinematic equations presented
in~\cite{Mikhailov-etal-1994} showed attachment of spiral waves to
non-flux boundaries, which in a sense is similar to attachment to
stepwise inhomogeneity. On a phenomenogical level, such attachment
is, of course well known since the earliest simulations of excitable media,
\eg~\cite{Ermakova-Pertsov-1986}.

\paragraph{Orbiting drift vs other spiral wave dynamics.}
Properties of the orbital drift resemble properties of resonant drift
when the stimulation frequency is not fixed as in the examples above,
but is controlled by feed-back~\cite{Biktashev-Holden-1994}. In that
case, the dynamics of the spiral wave is controlled by a closed
autonomous system of two differential equations for the instant centre
of rotation of the spiral, like~\eq{stepODE} or \eq{diskODE}. In
particular, depending on the detail of the feedback, this planar
system may have limit cycle attractors, dubbed ``resonant attractors''
in~\cite{Zykov-Engel-2004}, which may have circular shape if
the system with the feedback has an axial symmetry. Apart from this
being a completely different type of drift, we also comment that the
second order ODE system is an approximation subject to the assumption
that the feedback is instant, and in the situations when the delay in
the feedback is significant due to the system size and large distance
between spiral core and feedback electrode, the behaviour becomes more
complicated.

For some combination of parameters, the trajectory of an orbiting
spiral may also resemble meandering and may be taken for this in
simulations or experiments. So, it is possible that orbital movement
was actually observed by Zou \etal\ \cite[p.802]{Zou-etal-1993} where
they reported spiral ``meandering'' around a ``partially excitable
defect''; although it is difficult to be certain as no details are
given.  The difference is that spiral meandering, in the proper sense,
is due to internal instabilities of a spiral wave, whereas orbital
motion is due to inhomogeneity. E.g. in orbiting, the ``meandering
pattern'' determined by $\Omega/\omega$ will change depending on the
inhomogeneity strength.

The phenomenon of ``pinning'' of spiral waves to localized
inhomogeneities has important practical implications for the problem
of low voltage defibrillation~\cite{%
  Krinsky-etal-1990,%
  Pumir-Krinsky-1999,%
  Pazo-etal-2004,%
  Ripplinger-etal-2006%
}. In terms of spiral wave
dynamics this is usually understood as attraction of the spiral centre
towards the inhomogeneity locus. Practically interesting cases of
pinning are usually associated with inexcitable obstacles, which are
not small perturbations and therefore not amenable to the asymptotic
theory considered in this study. However, the possibility of orbital
motion around a weak inhomogeneity suggests that a similar phenomenon
may be observed in strong inhomogeneities as well. This offers an
unexpected aspect on the problem of pinning. Instead of a simplistic
``binary" viewpoint, that a local inhomogeneity can either be
attractive, which is the case of pinning, or repelling, which is the
case of unpinning, there is actually a third possibility, which can in
fact be more prevalent than the first two, namely, that at some
initial conditions the spiral may orbit around one of a number of
circular orbits, regardless of whether or not it is attracted to the
center, which can be considered just as one of the orbits that happens
to have radius zero. That is, there is more than one way that a spiral
may be bound to inhomogeneities.

\paragraph{Conclusion.}
We have demonstrated that the asymptotic theory of spiral wave drift
in response to small perturbations, presented in
\cite{Biktashev-Holden-1995,Biktasheva-Biktashev-2003}, works well for
excitable media, described by FitzHugh-Nagumo and Barkley kinetics
models and gives accurate quantitative prediction of the drift for a
wide selection of perturbations.

The key objects of the asymptotic theory are the response functions,
\ie\ the critical eigenfunctions of the adjoint linearized operator.
The RFs have been found to be localized in most models
where they have been calculated; however there are counterexamples
demonstrating that surprises are possible~\cite{Biktashev-2005}.
Physical intuition tells that for the response function to be
localized, the spiral wave should be indifferent to distant
perturbations, which will be the case if the core of the spiral is a
``source'' rather than a ``sink'' in the sense of the flow of
causality, for example as defined by the group velocity. Indeed, this
localization property has been proven for one-dimensional analogues of
spiral waves~\cite{Sandstede-Scheel-2004} and there is hope that this
result can be extended to spiral and scroll waves.

The effective spatial localization of the RFs on the mathematical
level guarantees convergence of the integrals involved in asymptotic
theory, and on the physical level explains why wave-like objects like
spiral and scroll waves, while stretching to infinity and
synchronizing the whole medium, behave respectively as particle-like
and string-like localized objects. This macroscopic dissipative
wave-particle duality of the spiral waves has been previously
demonstrated for the Complex Ginzburg-Landau equation
\cite{Biktasheva-Biktashev-2003} which is the archetypical oscillatory
media model. Here we confirmed it for the most popular excitable media
models important for many applications.

\section*{Acknowledgement}
This study has been supported in part by EPSRC grants EP/D074789/1 and
EP/D074746/1. DB also gratefully acknowledges support from
  the Leverhulme Trust and the Royal Society.


\begin{thebibliography}{67}
\expandafter\ifx\csname natexlab\endcsname\relax\def\natexlab#1{#1}\fi
\expandafter\ifx\csname bibnamefont\endcsname\relax
  \def\bibnamefont#1{#1}\fi
\expandafter\ifx\csname bibfnamefont\endcsname\relax
  \def\bibfnamefont#1{#1}\fi
\expandafter\ifx\csname citenamefont\endcsname\relax
  \def\citenamefont#1{#1}\fi
\expandafter\ifx\csname url\endcsname\relax
  \def\url#1{\texttt{#1}}\fi
\expandafter\ifx\csname urlprefix\endcsname\relax\def\urlprefix{URL }\fi
\providecommand{\bibinfo}[2]{#2}
\providecommand{\eprint}[2][]{\url{#2}}

\bibitem[{\citenamefont{Frisch et~al.}(1994)\citenamefont{Frisch, Rica,
  Coullet, and Gilli}}]{Frisch-etal-1994}
\bibinfo{author}{\bibfnamefont{T.}~\bibnamefont{Frisch}},
  \bibinfo{author}{\bibfnamefont{S.}~\bibnamefont{Rica}},
  \bibinfo{author}{\bibfnamefont{P.}~\bibnamefont{Coullet}}, \bibnamefont{and}
  \bibinfo{author}{\bibfnamefont{J.~M.} \bibnamefont{Gilli}},
  \bibinfo{journal}{Phys. Rev. Lett.} \textbf{\bibinfo{volume}{72}},
  \bibinfo{pages}{1471} (\bibinfo{year}{1994}).

\bibitem[{\citenamefont{Madore and Freedman}(1987)}]{Madore-Freedman-1987}
\bibinfo{author}{\bibfnamefont{B.~F.} \bibnamefont{Madore}} \bibnamefont{and}
  \bibinfo{author}{\bibfnamefont{W.~L.} \bibnamefont{Freedman}},
  \bibinfo{journal}{Am. Sci.} \textbf{\bibinfo{volume}{75}},
  \bibinfo{pages}{252} (\bibinfo{year}{1987}).

\bibitem[{\citenamefont{Schulman and Seiden}(1986)}]{Schulman-Seiden-1986}
\bibinfo{author}{\bibfnamefont{L.~S.} \bibnamefont{Schulman}} \bibnamefont{and}
  \bibinfo{author}{\bibfnamefont{P.~E.} \bibnamefont{Seiden}},
  \bibinfo{journal}{Science} \textbf{\bibinfo{volume}{233}},
  \bibinfo{pages}{425} (\bibinfo{year}{1986}).

\bibitem[{\citenamefont{Zhabotinsky and
  Zaikin}(1971)}]{Zhabotinsky-Zaikin-1971}
\bibinfo{author}{\bibfnamefont{A.~M.} \bibnamefont{Zhabotinsky}}
  \bibnamefont{and} \bibinfo{author}{\bibfnamefont{A.~N.}
  \bibnamefont{Zaikin}}, in \emph{\bibinfo{booktitle}{Oscillatory processes in
  biological and chemical systems}}, edited by
  \bibinfo{editor}{\bibfnamefont{E.~E.} \bibnamefont{Selkov}},
  \bibinfo{editor}{\bibfnamefont{A.~A.} \bibnamefont{Zhabotinsky}},
  \bibnamefont{and} \bibinfo{editor}{\bibfnamefont{S.~E.} \bibnamefont{Shnol}}
  (\bibinfo{publisher}{Nauka}, \bibinfo{address}{Pushchino},
  \bibinfo{year}{1971}), p. \bibinfo{pages}{279}, \bibinfo{note}{in Russian}.

\bibitem[{\citenamefont{Jakubith et~al.}(1990)\citenamefont{Jakubith,
  Rotermund, Engel, von Oertzen, and Ertl}}]{Jakubith-etal-1990}
\bibinfo{author}{\bibfnamefont{S.}~\bibnamefont{Jakubith}},
  \bibinfo{author}{\bibfnamefont{H.~H.} \bibnamefont{Rotermund}},
  \bibinfo{author}{\bibfnamefont{W.}~\bibnamefont{Engel}},
  \bibinfo{author}{\bibfnamefont{A.}~\bibnamefont{von Oertzen}},
  \bibnamefont{and} \bibinfo{author}{\bibfnamefont{G.}~\bibnamefont{Ertl}},
  \bibinfo{journal}{Phys. Rev. Lett.} \textbf{\bibinfo{volume}{65}},
  \bibinfo{pages}{3013} (\bibinfo{year}{1990}).

\bibitem[{\citenamefont{Allessie et~al.}(1973)\citenamefont{Allessie, Bonke,
  and Schopman}}]{Allessie-etal-1973}
\bibinfo{author}{\bibfnamefont{M.~A.} \bibnamefont{Allessie}},
  \bibinfo{author}{\bibfnamefont{F.~I.~M.} \bibnamefont{Bonke}},
  \bibnamefont{and} \bibinfo{author}{\bibfnamefont{F.~J.~G.}
  \bibnamefont{Schopman}}, \bibinfo{journal}{Circ. Res.}
  \textbf{\bibinfo{volume}{33}}, \bibinfo{pages}{54} (\bibinfo{year}{1973}).

\bibitem[{\citenamefont{Gorelova and Bures}(1983)}]{Gorelova-Bures-1983}
\bibinfo{author}{\bibfnamefont{N.~A.} \bibnamefont{Gorelova}} \bibnamefont{and}
  \bibinfo{author}{\bibfnamefont{J.}~\bibnamefont{Bures}}, \bibinfo{journal}{J.
  Neurobiol.} \textbf{\bibinfo{volume}{14}}, \bibinfo{pages}{353}
  (\bibinfo{year}{1983}).

\bibitem[{\citenamefont{Alcantara and Monk}(1974)}]{Alcantara-Monk-1974}
\bibinfo{author}{\bibfnamefont{F.}~\bibnamefont{Alcantara}} \bibnamefont{and}
  \bibinfo{author}{\bibfnamefont{M.}~\bibnamefont{Monk}}, \bibinfo{journal}{J.
  Gen. Microbiol.} \textbf{\bibinfo{volume}{85}}, \bibinfo{pages}{321}
  (\bibinfo{year}{1974}).

\bibitem[{\citenamefont{Lechleiter et~al.}(1991)\citenamefont{Lechleiter,
  Girard, Peralta, and Clapham}}]{Lechleiter-etal-1991}
\bibinfo{author}{\bibfnamefont{J.}~\bibnamefont{Lechleiter}},
  \bibinfo{author}{\bibfnamefont{S.}~\bibnamefont{Girard}},
  \bibinfo{author}{\bibfnamefont{E.}~\bibnamefont{Peralta}}, \bibnamefont{and}
  \bibinfo{author}{\bibfnamefont{D.}~\bibnamefont{Clapham}},
  \bibinfo{journal}{Science} \textbf{\bibinfo{volume}{252}},
  \bibinfo{pages}{123} (\bibinfo{year}{1991}).

\bibitem[{\citenamefont{Carey et~al.}(1978)\citenamefont{Carey, Giles, and
  McLean}}]{Carey-etal-1978}
\bibinfo{author}{\bibfnamefont{A.~B.} \bibnamefont{Carey}},
  \bibinfo{author}{\bibfnamefont{R.~H.} \bibnamefont{Giles},
  \bibfnamefont{Jr.}}, \bibnamefont{and} \bibinfo{author}{\bibfnamefont{R.~G.}
  \bibnamefont{McLean}}, \bibinfo{journal}{Am. J. Trop. Med. Hyg.}
  \textbf{\bibinfo{volume}{27}}, \bibinfo{pages}{573} (\bibinfo{year}{1978}).

\bibitem[{\citenamefont{Murray et~al.}(1986)\citenamefont{Murray, Stanley, and
  Brown}}]{Murray-etal-1986}
\bibinfo{author}{\bibfnamefont{J.~D.} \bibnamefont{Murray}},
  \bibinfo{author}{\bibfnamefont{E.~A.} \bibnamefont{Stanley}},
  \bibnamefont{and} \bibinfo{author}{\bibfnamefont{D.~L.} \bibnamefont{Brown}},
  \bibinfo{journal}{Proc. Roy. Soc. Lond. ser. B}
  \textbf{\bibinfo{volume}{229}}, \bibinfo{pages}{111} (\bibinfo{year}{1986}).

\bibitem[{\citenamefont{Shagalov}(1997)}]{Shagalov-1997}
\bibinfo{author}{\bibfnamefont{A.~G.} \bibnamefont{Shagalov}},
  \bibinfo{journal}{Phys. Lett. A} \textbf{\bibinfo{volume}{235}},
  \bibinfo{pages}{643} (\bibinfo{year}{1997}).

\bibitem[{\citenamefont{Oswald and Dequidt}(2008)}]{Oswald-Dequidt-2008}
\bibinfo{author}{\bibfnamefont{P.}~\bibnamefont{Oswald}} \bibnamefont{and}
  \bibinfo{author}{\bibfnamefont{A.}~\bibnamefont{Dequidt}},
  \bibinfo{journal}{Phys. Rev. E} \textbf{\bibinfo{volume}{77}},
  \bibinfo{pages}{051706} (\bibinfo{year}{2008}).

\bibitem[{\citenamefont{Larionova et~al.}(2005)\citenamefont{Larionova, Egorov,
  Cabrera-Granado, and Esteban-Martin}}]{Larionova-etal-2005}
\bibinfo{author}{\bibfnamefont{Y.}~\bibnamefont{Larionova}},
  \bibinfo{author}{\bibfnamefont{O.}~\bibnamefont{Egorov}},
  \bibinfo{author}{\bibfnamefont{E.}~\bibnamefont{Cabrera-Granado}},
  \bibnamefont{and}
  \bibinfo{author}{\bibfnamefont{A.}~\bibnamefont{Esteban-Martin}},
  \bibinfo{journal}{Phys. Rev. A} \textbf{\bibinfo{volume}{72}},
  \bibinfo{pages}{033825} (\bibinfo{year}{2005}).

\bibitem[{\citenamefont{Yu et~al.}(1999)\citenamefont{Yu, Lu, and
  Harrison}}]{Yu-etal-1999}
\bibinfo{author}{\bibfnamefont{D.~J.} \bibnamefont{Yu}},
  \bibinfo{author}{\bibfnamefont{W.~P.} \bibnamefont{Lu}}, \bibnamefont{and}
  \bibinfo{author}{\bibfnamefont{R.~G.} \bibnamefont{Harrison}},
  \bibinfo{journal}{Journal of Optics B --- Quantum and Semiclassical Optics}
  \textbf{\bibinfo{volume}{1}}, \bibinfo{pages}{25} (\bibinfo{year}{1999}).

\bibitem[{\citenamefont{Agladze and Steinbock}(2000)}]{Agladze-Steinbock-2000}
\bibinfo{author}{\bibfnamefont{K.}~\bibnamefont{Agladze}} \bibnamefont{and}
  \bibinfo{author}{\bibfnamefont{O.}~\bibnamefont{Steinbock}},
  \bibinfo{journal}{J.Phys.Chem. A} \textbf{\bibinfo{volume}{104}},
  \bibinfo{pages}{9816} (\bibinfo{year}{2000}).

\bibitem[{\citenamefont{Bretschneider et~al.}(2009)\citenamefont{Bretschneider,
  Anderson, Ecke, M{\"u}ller-Taubenberger, Schroth-Diez, Ishikawa-Ankerhold,
  and Gerisch}}]{Bretschneider-etal-2009}
\bibinfo{author}{\bibfnamefont{T.}~\bibnamefont{Bretschneider}},
  \bibinfo{author}{\bibfnamefont{K.}~\bibnamefont{Anderson}},
  \bibinfo{author}{\bibfnamefont{M.}~\bibnamefont{Ecke}},
  \bibinfo{author}{\bibfnamefont{A.}~\bibnamefont{M{\"u}ller-Taubenberger}},
  \bibinfo{author}{\bibfnamefont{B.}~\bibnamefont{Schroth-Diez}},
  \bibinfo{author}{\bibfnamefont{H.~C.} \bibnamefont{Ishikawa-Ankerhold}},
  \bibnamefont{and} \bibinfo{author}{\bibfnamefont{G.}~\bibnamefont{Gerisch}},
  \bibinfo{journal}{Biophys. J.} \textbf{\bibinfo{volume}{96}},
  \bibinfo{pages}{2888} (\bibinfo{year}{2009}).

\bibitem[{\citenamefont{Dahlem and M{\"u}ller}(2003)}]{Dahlem-Mueller-2003}
\bibinfo{author}{\bibfnamefont{M.~A.} \bibnamefont{Dahlem}} \bibnamefont{and}
  \bibinfo{author}{\bibfnamefont{S.~C.} \bibnamefont{M{\"u}ller}},
  \bibinfo{journal}{Biological Cybernetics} \textbf{\bibinfo{volume}{88}},
  \bibinfo{pages}{419} (\bibinfo{year}{2003}).

\bibitem[{\citenamefont{Igoshin et~al.}(2004)\citenamefont{Igoshin, Welch,
  Kaiser, and Oster}}]{Igoshin-etal-2004}
\bibinfo{author}{\bibfnamefont{O.~A.} \bibnamefont{Igoshin}},
  \bibinfo{author}{\bibfnamefont{R.}~\bibnamefont{Welch}},
  \bibinfo{author}{\bibfnamefont{D.}~\bibnamefont{Kaiser}}, \bibnamefont{and}
  \bibinfo{author}{\bibfnamefont{G.}~\bibnamefont{Oster}},
  \bibinfo{journal}{Proc. Nat. Acad. Sci. USA} \textbf{\bibinfo{volume}{101}},
  \bibinfo{pages}{4256} (\bibinfo{year}{2004}).

\bibitem[{\citenamefont{Agladze et~al.}(1987)\citenamefont{Agladze, Davydov,
  and Mikhailov}}]{Agladze-etal-1987}
\bibinfo{author}{\bibfnamefont{K.~I.} \bibnamefont{Agladze}},
  \bibinfo{author}{\bibfnamefont{V.~A.} \bibnamefont{Davydov}},
  \bibnamefont{and} \bibinfo{author}{\bibfnamefont{A.~S.}
  \bibnamefont{Mikhailov}}, \bibinfo{journal}{Letters to ZhETPh}
  \textbf{\bibinfo{volume}{45}}, \bibinfo{pages}{601} (\bibinfo{year}{1987}),
  \bibinfo{note}{in Russian}.

\bibitem[{\citenamefont{Agladze and Dekepper}(1992)}]{Agladze-Dekepper-1992}
\bibinfo{author}{\bibfnamefont{K.~I.} \bibnamefont{Agladze}} \bibnamefont{and}
  \bibinfo{author}{\bibfnamefont{P.}~\bibnamefont{Dekepper}},
  \bibinfo{journal}{Journal of Physical Chemistry}
  \textbf{\bibinfo{volume}{96}}, \bibinfo{pages}{5239} (\bibinfo{year}{1992}).

\bibitem[{\citenamefont{Fast and Pertsov}(1992)}]{Fast-Pertsov-1992}
\bibinfo{author}{\bibfnamefont{V.~G.} \bibnamefont{Fast}} \bibnamefont{and}
  \bibinfo{author}{\bibfnamefont{A.~M.} \bibnamefont{Pertsov}},
  \bibinfo{journal}{J. Cardiovasc. Electrophysiol.}
  \textbf{\bibinfo{volume}{3}}, \bibinfo{pages}{255} (\bibinfo{year}{1992}).

\bibitem[{\citenamefont{Davidenko et~al.}(1992)\citenamefont{Davidenko,
  Pertsov, Salamonsz, Baxter, and Jalife}}]{Davidenko-etal-1992}
\bibinfo{author}{\bibfnamefont{J.~M.} \bibnamefont{Davidenko}},
  \bibinfo{author}{\bibfnamefont{A.~M.} \bibnamefont{Pertsov}},
  \bibinfo{author}{\bibfnamefont{R.}~\bibnamefont{Salamonsz}},
  \bibinfo{author}{\bibfnamefont{W.}~\bibnamefont{Baxter}}, \bibnamefont{and}
  \bibinfo{author}{\bibfnamefont{J.}~\bibnamefont{Jalife}},
  \bibinfo{journal}{Nature} \textbf{\bibinfo{volume}{335}},
  \bibinfo{pages}{349} (\bibinfo{year}{1992}).

\bibitem[{\citenamefont{Markus et~al.}(1992)\citenamefont{Markus,
  Nagy-Ungvarai, and Hess}}]{Markus-etal-1992}
\bibinfo{author}{\bibfnamefont{M.}~\bibnamefont{Markus}},
  \bibinfo{author}{\bibfnamefont{Z.}~\bibnamefont{Nagy-Ungvarai}},
  \bibnamefont{and} \bibinfo{author}{\bibfnamefont{B.}~\bibnamefont{Hess}},
  \bibinfo{journal}{Science} \textbf{\bibinfo{volume}{257}},
  \bibinfo{pages}{225} (\bibinfo{year}{1992}).

\bibitem[{\citenamefont{Luengviriya et~al.}(2006)\citenamefont{Luengviriya,
  Storb, Hauser, and C.}}]{Luengviriya-etal-2006}
\bibinfo{author}{\bibfnamefont{C.}~\bibnamefont{Luengviriya}},
  \bibinfo{author}{\bibfnamefont{U.}~\bibnamefont{Storb}},
  \bibinfo{author}{\bibfnamefont{M.~J.~B.} \bibnamefont{Hauser}},
  \bibnamefont{and} \bibinfo{author}{\bibfnamefont{M.~S.} \bibnamefont{C.}},
  \bibinfo{journal}{Phys. Chem. Chem. Phys.} \textbf{\bibinfo{volume}{8}},
  \bibinfo{pages}{1425} (\bibinfo{year}{2006}).

\bibitem[{\citenamefont{Nettesheim et~al.}(1993)\citenamefont{Nettesheim, von
  Oertzen, Rotermund, and Ertl}}]{Nettesheim-etal-1993}
\bibinfo{author}{\bibfnamefont{S.}~\bibnamefont{Nettesheim}},
  \bibinfo{author}{\bibfnamefont{A.}~\bibnamefont{von Oertzen}},
  \bibinfo{author}{\bibfnamefont{H.~H.} \bibnamefont{Rotermund}},
  \bibnamefont{and} \bibinfo{author}{\bibfnamefont{G.}~\bibnamefont{Ertl}},
  \bibinfo{journal}{J. Chem. Phys.} \textbf{\bibinfo{volume}{98}},
  \bibinfo{pages}{9977} (\bibinfo{year}{1993}).

\bibitem[{\citenamefont{Pertsov et~al.}(1993)\citenamefont{Pertsov, Davidenko,
  Salomonsz, Baxter, and Jalife}}]{Pertsov-etal-1993}
\bibinfo{author}{\bibfnamefont{A.~M.} \bibnamefont{Pertsov}},
  \bibinfo{author}{\bibfnamefont{J.~M.} \bibnamefont{Davidenko}},
  \bibinfo{author}{\bibfnamefont{R.}~\bibnamefont{Salomonsz}},
  \bibinfo{author}{\bibfnamefont{W.~T.} \bibnamefont{Baxter}},
  \bibnamefont{and} \bibinfo{author}{\bibfnamefont{J.}~\bibnamefont{Jalife}},
  \bibinfo{journal}{Circ. Res.} \textbf{\bibinfo{volume}{72}},
  \bibinfo{pages}{631} (\bibinfo{year}{1993}).

\bibitem[{\citenamefont{Lim et~al.}(2006)\citenamefont{Lim, Maskara, Aguel,
  Emokpae, and Tung}}]{Lim-etal-2006}
\bibinfo{author}{\bibfnamefont{Z.~Y.} \bibnamefont{Lim}},
  \bibinfo{author}{\bibfnamefont{B.}~\bibnamefont{Maskara}},
  \bibinfo{author}{\bibfnamefont{F.}~\bibnamefont{Aguel}},
  \bibinfo{author}{\bibfnamefont{R.}~\bibnamefont{Emokpae}}, \bibnamefont{and}
  \bibinfo{author}{\bibfnamefont{L.}~\bibnamefont{Tung}},
  \bibinfo{journal}{Circulation} \textbf{\bibinfo{volume}{114}},
  \bibinfo{pages}{2113} (\bibinfo{year}{2006}).

\bibitem[{\citenamefont{Lugomer et~al.}(2007)\citenamefont{Lugomer, Fukumoto,
  Farkas, Sz{\"or\'e}nyi, and Toth}}]{Lugomer-etal-2007}
\bibinfo{author}{\bibfnamefont{S.}~\bibnamefont{Lugomer}},
  \bibinfo{author}{\bibfnamefont{Y.}~\bibnamefont{Fukumoto}},
  \bibinfo{author}{\bibfnamefont{B.}~\bibnamefont{Farkas}},
  \bibinfo{author}{\bibfnamefont{T.}~\bibnamefont{Sz{\"or\'e}nyi}},
  \bibnamefont{and} \bibinfo{author}{\bibfnamefont{A.}~\bibnamefont{Toth}},
  \bibinfo{journal}{Phys. Rev. E} \textbf{\bibinfo{volume}{76}},
  \bibinfo{pages}{016305} (\bibinfo{year}{2007}).

\bibitem[{\citenamefont{Yakushevich}(1984)}]{Yakushevich-1984}
\bibinfo{author}{\bibfnamefont{L.}~\bibnamefont{Yakushevich}},
  \bibinfo{journal}{Studia Biophysica} \textbf{\bibinfo{volume}{100}},
  \bibinfo{pages}{195} (\bibinfo{year}{1984}).

\bibitem[{\citenamefont{Pertsov and Ermakova}(1988)}]{Pertsov-Ermakova-1988}
\bibinfo{author}{\bibfnamefont{A.~M.} \bibnamefont{Pertsov}} \bibnamefont{and}
  \bibinfo{author}{\bibfnamefont{E.~A.} \bibnamefont{Ermakova}},
  \bibinfo{journal}{Biofizika} \textbf{\bibinfo{volume}{33}},
  \bibinfo{pages}{338} (\bibinfo{year}{1988}), \bibinfo{note}{in Russian}.

\bibitem[{\citenamefont{Wellner et~al.}(1999)\citenamefont{Wellner, Pertsov,
  and Jalife}}]{Wellner-etal-1999}
\bibinfo{author}{\bibfnamefont{M.}~\bibnamefont{Wellner}},
  \bibinfo{author}{\bibfnamefont{A.~M.} \bibnamefont{Pertsov}},
  \bibnamefont{and} \bibinfo{author}{\bibfnamefont{J.}~\bibnamefont{Jalife}},
  \bibinfo{journal}{Phys. Rev. E} \textbf{\bibinfo{volume}{59}},
  \bibinfo{pages}{5192} (\bibinfo{year}{1999}).

\bibitem[{\citenamefont{Hendrey et~al.}(2000)\citenamefont{Hendrey, Ott, and
  Antonsen}}]{Hendrey-etal-2000}
\bibinfo{author}{\bibfnamefont{M.}~\bibnamefont{Hendrey}},
  \bibinfo{author}{\bibfnamefont{E.}~\bibnamefont{Ott}}, \bibnamefont{and}
  \bibinfo{author}{\bibfnamefont{T.~M.} \bibnamefont{Antonsen}},
  \bibinfo{journal}{Phys. Rev. E} \textbf{\bibinfo{volume}{61}},
  \bibinfo{pages}{4943} (\bibinfo{year}{2000}).

\bibitem[{\citenamefont{Keener}(1988)}]{Keener-1988}
\bibinfo{author}{\bibfnamefont{J.~P.} \bibnamefont{Keener}},
  \bibinfo{journal}{Physica D} \textbf{\bibinfo{volume}{31}},
  \bibinfo{pages}{269} (\bibinfo{year}{1988}).

\bibitem[{\citenamefont{Biktashev et~al.}(1994)\citenamefont{Biktashev, Holden,
  and Zhang}}]{Biktashev-etal-1994}
\bibinfo{author}{\bibfnamefont{V.~N.} \bibnamefont{Biktashev}},
  \bibinfo{author}{\bibfnamefont{A.~V.} \bibnamefont{Holden}},
  \bibnamefont{and} \bibinfo{author}{\bibfnamefont{H.}~\bibnamefont{Zhang}},
  \bibinfo{journal}{Phil. Trans. Roy. Soc. Lond. ser. A}
  \textbf{\bibinfo{volume}{347}}, \bibinfo{pages}{611} (\bibinfo{year}{1994}).

\bibitem[{\citenamefont{Biktashev and Holden}(1995)}]{Biktashev-Holden-1995}
\bibinfo{author}{\bibfnamefont{V.~N.} \bibnamefont{Biktashev}}
  \bibnamefont{and} \bibinfo{author}{\bibfnamefont{A.~V.}
  \bibnamefont{Holden}}, \bibinfo{journal}{Chaos, Solitons and Fractals}
  \textbf{\bibinfo{volume}{5}}, \bibinfo{pages}{575} (\bibinfo{year}{1995}).

\bibitem[{\citenamefont{Biktashev}(1989)}]{Biktashev-1989}
\bibinfo{author}{\bibfnamefont{V.~N.} \bibnamefont{Biktashev}}, Ph.D. thesis,
  \bibinfo{school}{MPhTI} (\bibinfo{year}{1989}), \bibinfo{note}{in Russian}.

\bibitem[{\citenamefont{Ermakova and Pertsov}(1986)}]{Ermakova-Pertsov-1986}
\bibinfo{author}{\bibfnamefont{E.~A.} \bibnamefont{Ermakova}} \bibnamefont{and}
  \bibinfo{author}{\bibfnamefont{A.~N.} \bibnamefont{Pertsov}},
  \bibinfo{journal}{Biofizika} \textbf{\bibinfo{volume}{31}},
  \bibinfo{pages}{855} (\bibinfo{year}{1986}), \bibinfo{note}{in Russian}.

\bibitem[{\citenamefont{Ermakova et~al.}(1987)\citenamefont{Ermakova, Pertsov,
  and Shnol}}]{Ermakova-etal-1987}
\bibinfo{author}{\bibfnamefont{E.~A.} \bibnamefont{Ermakova}},
  \bibinfo{author}{\bibfnamefont{A.~N.} \bibnamefont{Pertsov}},
  \bibnamefont{and} \bibinfo{author}{\bibfnamefont{E.~E.} \bibnamefont{Shnol}},
  \emph{\bibinfo{title}{Couples of the interacting vortices in two-dimensional
  active media}} (\bibinfo{publisher}{ONTI NCBI}, \bibinfo{address}{Pushchino},
  \bibinfo{year}{1987}), \bibinfo{note}{in Russian}.

\bibitem[{\citenamefont{Ermakova et~al.}(1989)\citenamefont{Ermakova, Pertsov,
  and Shnol}}]{Ermakova-etal-1989}
\bibinfo{author}{\bibfnamefont{E.~A.} \bibnamefont{Ermakova}},
  \bibinfo{author}{\bibfnamefont{A.~M.} \bibnamefont{Pertsov}},
  \bibnamefont{and} \bibinfo{author}{\bibfnamefont{E.~E.} \bibnamefont{Shnol}},
  \bibinfo{journal}{Physica D} \textbf{\bibinfo{volume}{40}},
  \bibinfo{pages}{185} (\bibinfo{year}{1989}).

\bibitem[{\citenamefont{Hamm}(1997)}]{Hamm-1997}
\bibinfo{author}{\bibfnamefont{E.}~\bibnamefont{Hamm}}, Ph.D. thesis,
  \bibinfo{school}{Universit\'e de Nice - Sophia Antipolice / Institut Non
  Lin\'eair de Nice} (\bibinfo{year}{1997}), \bibinfo{note}{in French}.

\bibitem[{\citenamefont{Biktasheva et~al.}(1998)\citenamefont{Biktasheva,
  Elkin, and Biktashev}}]{Biktasheva-etal-1998}
\bibinfo{author}{\bibfnamefont{I.~V.} \bibnamefont{Biktasheva}},
  \bibinfo{author}{\bibfnamefont{Y.~E.} \bibnamefont{Elkin}}, \bibnamefont{and}
  \bibinfo{author}{\bibfnamefont{V.~N.} \bibnamefont{Biktashev}},
  \bibinfo{journal}{Phys. Rev. E} \textbf{\bibinfo{volume}{57}},
  \bibinfo{pages}{2656} (\bibinfo{year}{1998}).

\bibitem[{\citenamefont{Biktasheva et~al.}(1999)\citenamefont{Biktasheva,
  Elkin, and Biktashev}}]{Biktasheva-etal-1999}
\bibinfo{author}{\bibfnamefont{I.~V.} \bibnamefont{Biktasheva}},
  \bibinfo{author}{\bibfnamefont{Y.~E.} \bibnamefont{Elkin}}, \bibnamefont{and}
  \bibinfo{author}{\bibfnamefont{V.~N.} \bibnamefont{Biktashev}},
  \bibinfo{journal}{J. Biol. Phys.} \textbf{\bibinfo{volume}{25}},
  \bibinfo{pages}{115} (\bibinfo{year}{1999}).

\bibitem[{\citenamefont{Biktasheva}(2000{\natexlab{a}})}]{Biktasheva-2000}
\bibinfo{author}{\bibfnamefont{I.~V.} \bibnamefont{Biktasheva}},
  \bibinfo{journal}{Phys. Rev. E} \textbf{\bibinfo{volume}{62}},
  \bibinfo{pages}{8800} (\bibinfo{year}{2000}{\natexlab{a}}).

\bibitem[{\citenamefont{Biktasheva and
  Biktashev}(2003)}]{Biktasheva-Biktashev-2003}
\bibinfo{author}{\bibfnamefont{I.~V.} \bibnamefont{Biktasheva}}
  \bibnamefont{and} \bibinfo{author}{\bibfnamefont{V.~N.}
  \bibnamefont{Biktashev}}, \bibinfo{journal}{Phys. Rev. E}
  \textbf{\bibinfo{volume}{67}}, \bibinfo{pages}{026221}
  (\bibinfo{year}{2003}).

\bibitem[{\citenamefont{Henry and Hakim}(2002)}]{Henry-Hakim-2002}
\bibinfo{author}{\bibfnamefont{H.}~\bibnamefont{Henry}} \bibnamefont{and}
  \bibinfo{author}{\bibfnamefont{V.}~\bibnamefont{Hakim}},
  \bibinfo{journal}{Phys. Rev. E} \textbf{\bibinfo{volume}{65}},
  \bibinfo{pages}{046235} (\bibinfo{year}{2002}).

\bibitem[{\citenamefont{Biktasheva et~al.}(2009)\citenamefont{Biktasheva,
  Barkley, Biktashev, Bordyugov, and Foulkes}}]{Biktasheva-etal-2009}
\bibinfo{author}{\bibfnamefont{I.~V.} \bibnamefont{Biktasheva}},
  \bibinfo{author}{\bibfnamefont{D.}~\bibnamefont{Barkley}},
  \bibinfo{author}{\bibfnamefont{V.~N.} \bibnamefont{Biktashev}},
  \bibinfo{author}{\bibfnamefont{G.~V.} \bibnamefont{Bordyugov}},
  \bibnamefont{and} \bibinfo{author}{\bibfnamefont{A.~J.}
  \bibnamefont{Foulkes}}, \bibinfo{journal}{Phys. Rev. E}
  \textbf{\bibinfo{volume}{79}}, \bibinfo{pages}{056702}
  (\bibinfo{year}{2009}).

\bibitem[{\citenamefont{FitzHugh}(1961)}]{FitzHugh-1961}
\bibinfo{author}{\bibfnamefont{R.}~\bibnamefont{FitzHugh}},
  \bibinfo{journal}{Biophys. J.} \textbf{\bibinfo{volume}{1}},
  \bibinfo{pages}{445} (\bibinfo{year}{1961}).

\bibitem[{\citenamefont{Nagumo et~al.}(1962)\citenamefont{Nagumo, Arimoto, and
  Yoshizawa}}]{Nagumo-etal-1962}
\bibinfo{author}{\bibfnamefont{J.~S.} \bibnamefont{Nagumo}},
  \bibinfo{author}{\bibfnamefont{S.}~\bibnamefont{Arimoto}}, \bibnamefont{and}
  \bibinfo{author}{\bibfnamefont{S.}~\bibnamefont{Yoshizawa}},
  \bibinfo{journal}{Proc. IRE} \textbf{\bibinfo{volume}{50}},
  \bibinfo{pages}{2061} (\bibinfo{year}{1962}).

\bibitem[{\citenamefont{Winfree}(1991)}]{Winfree-1991}
\bibinfo{author}{\bibfnamefont{A.~T.} \bibnamefont{Winfree}},
  \bibinfo{journal}{Chaos} \textbf{\bibinfo{volume}{1}}, \bibinfo{pages}{303}
  (\bibinfo{year}{1991}).

\bibitem[{\citenamefont{Barkley}(1991)}]{Barkley-1991}
\bibinfo{author}{\bibfnamefont{D.}~\bibnamefont{Barkley}},
  \bibinfo{journal}{Physica D} \textbf{\bibinfo{volume}{49}},
  \bibinfo{pages}{61} (\bibinfo{year}{1991}).

\bibitem[{\citenamefont{Biktashev et~al.}(2010)\citenamefont{Biktashev,
  Barkley, and Biktasheva}}]{OMS}
\bibinfo{author}{\bibfnamefont{V.~N.} \bibnamefont{Biktashev}},
  \bibinfo{author}{\bibfnamefont{D.}~\bibnamefont{Barkley}}, \bibnamefont{and}
  \bibinfo{author}{\bibfnamefont{I.~V.} \bibnamefont{Biktasheva}},
  \bibinfo{journal}{Phys. Rev. Lett.} \textbf{\bibinfo{volume}{104}},
  \bibinfo{pages}{058302} (\bibinfo{year}{2010}).

\bibitem[{\citenamefont{LeBlanc and Wulff}(2000)}]{LeBlanc-Wulff-2000}
\bibinfo{author}{\bibfnamefont{V.~G.} \bibnamefont{LeBlanc}} \bibnamefont{and}
  \bibinfo{author}{\bibfnamefont{C.}~\bibnamefont{Wulff}}, \bibinfo{journal}{J.
  Nonlinear Sci.} \textbf{\bibinfo{volume}{10}}, \bibinfo{pages}{569}
  (\bibinfo{year}{2000}).

\bibitem[{\citenamefont{Biktasheva}(2000{\natexlab{b}})}]{Biktasheva-PhD-Russ}
\bibinfo{author}{\bibfnamefont{I.~V.} \bibnamefont{Biktasheva}}, Ph.D. thesis,
  \bibinfo{school}{ITEB RAS} (\bibinfo{year}{2000}{\natexlab{b}}),
  \bibinfo{note}{in Russian}.

\bibitem[{\citenamefont{Biktasheva and
  Biktashev}(2001)}]{Biktasheva-Biktashev-2001}
\bibinfo{author}{\bibfnamefont{I.~V.} \bibnamefont{Biktasheva}}
  \bibnamefont{and} \bibinfo{author}{\bibfnamefont{V.~N.}
  \bibnamefont{Biktashev}}, \bibinfo{journal}{J. Nonlin. Math. Phys.}
  \textbf{\bibinfo{volume}{8 Supl.}}, \bibinfo{pages}{28}
  (\bibinfo{year}{2001}).

\bibitem[{\citenamefont{Elkin and Biktashev}(1999)}]{Elkin-Biktashev-1999}
\bibinfo{author}{\bibfnamefont{Y.~E.} \bibnamefont{Elkin}} \bibnamefont{and}
  \bibinfo{author}{\bibfnamefont{V.~N.} \bibnamefont{Biktashev}},
  \bibinfo{journal}{J. Biol. Phys.} \textbf{\bibinfo{volume}{25}},
  \bibinfo{pages}{129} (\bibinfo{year}{1999}).

\bibitem[{\citenamefont{Mikhailov et~al.}(1994)\citenamefont{Mikhailov,
  Davydov, and Zykov}}]{Mikhailov-etal-1994}
\bibinfo{author}{\bibfnamefont{A.~S.} \bibnamefont{Mikhailov}},
  \bibinfo{author}{\bibfnamefont{V.~A.} \bibnamefont{Davydov}},
  \bibnamefont{and} \bibinfo{author}{\bibfnamefont{V.~S.} \bibnamefont{Zykov}},
  \bibinfo{journal}{Physica D} \textbf{\bibinfo{volume}{70}},
  \bibinfo{pages}{1} (\bibinfo{year}{1994}).

\bibitem[{\citenamefont{Hakim and Karma}(1999)}]{Hakim-Karma-1999}
\bibinfo{author}{\bibfnamefont{V.}~\bibnamefont{Hakim}} \bibnamefont{and}
  \bibinfo{author}{\bibfnamefont{A.}~\bibnamefont{Karma}},
  \bibinfo{journal}{Phys. Rev. E} \textbf{\bibinfo{volume}{60}},
  \bibinfo{pages}{5073} (\bibinfo{year}{1999}).

\bibitem[{\citenamefont{Biktashev and Holden}(1994)}]{Biktashev-Holden-1994}
\bibinfo{author}{\bibfnamefont{V.~N.} \bibnamefont{Biktashev}}
  \bibnamefont{and} \bibinfo{author}{\bibfnamefont{A.~V.}
  \bibnamefont{Holden}}, \bibinfo{journal}{J. Theor. Biol.}
  \textbf{\bibinfo{volume}{169}}, \bibinfo{pages}{101} (\bibinfo{year}{1994}).

\bibitem[{\citenamefont{Zykov and Engel}(2004)}]{Zykov-Engel-2004}
\bibinfo{author}{\bibfnamefont{V.~S.} \bibnamefont{Zykov}} \bibnamefont{and}
  \bibinfo{author}{\bibfnamefont{H.}~\bibnamefont{Engel}},
  \bibinfo{journal}{Physica D} \textbf{\bibinfo{volume}{199}},
  \bibinfo{pages}{243} (\bibinfo{year}{2004}).

\bibitem[{\citenamefont{Zou et~al.}(1993)\citenamefont{Zou, Levine, and
  Kessler}}]{Zou-etal-1993}
\bibinfo{author}{\bibfnamefont{X.}~\bibnamefont{Zou}},
  \bibinfo{author}{\bibfnamefont{H.}~\bibnamefont{Levine}}, \bibnamefont{and}
  \bibinfo{author}{\bibfnamefont{D.~A.} \bibnamefont{Kessler}},
  \bibinfo{journal}{Phys. Rev. E} \textbf{\bibinfo{volume}{47}},
  \bibinfo{pages}{R800} (\bibinfo{year}{1993}).

\bibitem[{\citenamefont{Krinsky et~al.}(1990)\citenamefont{Krinsky, Pertsov,
  and Biktashev}}]{Krinsky-etal-1990}
\bibinfo{author}{\bibfnamefont{V.~I.} \bibnamefont{Krinsky}},
  \bibinfo{author}{\bibfnamefont{A.~M.} \bibnamefont{Pertsov}},
  \bibnamefont{and} \bibinfo{author}{\bibfnamefont{V.~N.}
  \bibnamefont{Biktashev}}, \bibinfo{journal}{Ann. N. Y. Acad. Sci.}
  \textbf{\bibinfo{volume}{591}}, \bibinfo{pages}{232} (\bibinfo{year}{1990}).

\bibitem[{\citenamefont{Pumir and Krinsky}(1999)}]{Pumir-Krinsky-1999}
\bibinfo{author}{\bibfnamefont{A.}~\bibnamefont{Pumir}} \bibnamefont{and}
  \bibinfo{author}{\bibfnamefont{V.}~\bibnamefont{Krinsky}},
  \bibinfo{journal}{J. Theor. Biol.} \textbf{\bibinfo{volume}{199}},
  \bibinfo{pages}{311} (\bibinfo{year}{1999}).

\bibitem[{\citenamefont{Paz{\'o} et~al.}(2004)\citenamefont{Paz{\'o}, Kramer,
  Pumir, Kanani, Efimov, and Krinsky}}]{Pazo-etal-2004}
\bibinfo{author}{\bibfnamefont{D.}~\bibnamefont{Paz{\'o}}},
  \bibinfo{author}{\bibfnamefont{L.}~\bibnamefont{Kramer}},
  \bibinfo{author}{\bibfnamefont{A.}~\bibnamefont{Pumir}},
  \bibinfo{author}{\bibfnamefont{S.}~\bibnamefont{Kanani}},
  \bibinfo{author}{\bibfnamefont{I.}~\bibnamefont{Efimov}}, \bibnamefont{and}
  \bibinfo{author}{\bibfnamefont{V.}~\bibnamefont{Krinsky}},
  \bibinfo{journal}{Phys. Rev. Lett.} \textbf{\bibinfo{volume}{93}},
  \bibinfo{pages}{168303} (\bibinfo{year}{2004}).

\bibitem[{\citenamefont{Ripplinger et~al.}(2006)\citenamefont{Ripplinger,
  Krinsky, Nikolski, and Efimov}}]{Ripplinger-etal-2006}
\bibinfo{author}{\bibfnamefont{C.~M.} \bibnamefont{Ripplinger}},
  \bibinfo{author}{\bibfnamefont{V.~I.} \bibnamefont{Krinsky}},
  \bibinfo{author}{\bibfnamefont{V.~P.} \bibnamefont{Nikolski}},
  \bibnamefont{and} \bibinfo{author}{\bibfnamefont{I.~R.}
  \bibnamefont{Efimov}}, \bibinfo{journal}{Am. J. Physiol. --- Heart and Circ.
  Physiol.} \textbf{\bibinfo{volume}{291}}, \bibinfo{pages}{H184}
  (\bibinfo{year}{2006}).

\bibitem[{\citenamefont{Biktashev}(2005)}]{Biktashev-2005}
\bibinfo{author}{\bibfnamefont{V.~N.} \bibnamefont{Biktashev}},
  \bibinfo{journal}{Phys. Rev. Lett.} \textbf{\bibinfo{volume}{95}},
  \bibinfo{pages}{084501} (\bibinfo{year}{2005}).

\bibitem[{\citenamefont{Sandstede and Scheel}(2004)}]{Sandstede-Scheel-2004}
\bibinfo{author}{\bibfnamefont{B.}~\bibnamefont{Sandstede}} \bibnamefont{and}
  \bibinfo{author}{\bibfnamefont{A.}~\bibnamefont{Scheel}},
  \bibinfo{journal}{SIAM J. Appl. Dyn. Syst.} \textbf{\bibinfo{volume}{3}},
  \bibinfo{pages}{1} (\bibinfo{year}{2004}), \bibinfo{note}{{Sandstede} reports
  that these {1D} results can be extended to spiral waves and the paper about
  that is in preparation}.

\end{thebibliography}

\end{document}